\documentclass[aps,prf,twocolumn,superscriptaddress]{revtex4-2}
\usepackage{graphicx}
\usepackage{float}
\usepackage{color}
\usepackage{amsmath}
\usepackage[colorlinks=true, citecolor=red, linkcolor=blue,urlcolor=blue ]{hyperref}
\usepackage{appendix}
\usepackage{mathtools}
\usepackage{booktabs}
\usepackage{txfonts}

\usepackage{layouts}

\newcommand{\ex}{\mathbf{e}_x}

\newcommand{\ez}{\mathbf{e}_z}

\newcommand{\er}{\mathbf{e}_r}

\newcommand{\ephi}{\mathbf{e}_\phi}

\newcommand{\fb}{\mathbf{f}}
\newcommand{\ub}{\mathbf{u}}
\newcommand{\Fb}{\mathbf{F}}
\newcommand{\Tb}{\mathbf{T}}
\newcommand{\Ub}{\mathbf{U}}
\newcommand{\Omb}{\mathbf{\Omega}}
\newcommand{\db}{\mathbf{d}}

\newcommand{\Ab}{\mathbf{A}}
\newcommand{\Bb}{\mathbf{B}}
\newcommand{\Db}{\mathbf{D}}

\newcommand{\xb}{\mathbf{x}}

\newcommand{\Sb}{\mathbf{S}}
\newcommand{\Cb}{\mathbf{C}}
\newcommand{\Ib}{\mathbf{I}}

\newcommand{\phih}{\hat{\phi}}

\usepackage{verbatim}

\usepackage{upgreek}

\newcommand{\um}{$\mathrm{\upmu m}$}

\newlength{\cutwidth}
\newlength{\cutheight}

\usepackage{calculator}

\begin{document}
	\title{Elastohydrodynamic synchronization of rotating bacterial flagella}
	
	\author{Maria T\u{a}tulea-Codrean}
	\affiliation{Department of Applied Mathematics and Theoretical Physics, University of Cambridge, Cambridge CB3 0WA, United Kingdom}
	\affiliation{Coll\`{e}ge de France, 11 place Marcelin Berthelot, 75005 Paris, France}
	\author{Eric Lauga}
	\email[]{e.lauga@damtp.cam.ac.uk}
	\affiliation{Department of Applied Mathematics and Theoretical Physics, University of Cambridge, Cambridge CB3 0WA, United Kingdom}
	
	\date{\today}
	
	\begin{abstract}
		
		To rotate continuously without jamming, the flagellar filaments of bacteria need to be locked in phase. While several models have been proposed for  eukaryotic flagella,  the synchronization of bacterial flagella is less well understood. Starting from a reduced model of flexible and hydrodynamically-coupled bacterial flagella, we rigorously coarse-grain the equations of motion using the method of multiple scales, and hence show that bacterial flagella generically synchronize  to zero phase difference via an elastohydrodynamic mechanism. Remarkably, the { far-field} rate of synchronization is maximised at an intermediate value of elastic compliance, with surprising implications for bacteria.
		
	\end{abstract} 
	
	\maketitle

	Motile eukaryotic cells propelled by slender flagella or cilia   have long been known to display a feature essential to locomotion: the 
	synchronization of their swimming appendages~\cite{gray_book,Machemer1972}. From nearby spermatozoa matching the beating pattern of  their  flexible  flagella, to  ciliary carpets that actively pump fluids in a coordinated fashion, the  synchronization of eukaryotic flagella and cilia is ubiquitous~\cite{bray_book,lauga_book}. Multiple physical factors can induce synchronization, including direct hydrodynamic interactions between the filaments~\cite{Brumley2014,Elgeti2013}, elastic coupling through intracellular features~\cite{Quaranta2015,Wan2016}, steric interactions~\cite{Chelakkot2021} and coupling through the motion of the cell body~\cite{Geyer2013,Bennett2013}.  The mechanisms for synchronization have been uncovered by theoretical studies~\cite{Gueron1997,Elfring2011_physfluids,Elfring2011_jfm,Olson2015,Goldstein2016,Guo2018,Chakrabarti2019,Man2020,Guo2021,Liao2021}, amongst which a popular minimal approach is to model the tips of eukaryotic flagella and cilia as particles undergoing periodic motion above a rigid surface representing the cell~\cite{Vilfan2006,Niedermayer2008,Brumley2012,Uchida2011,Uchida2012,Tanasijevic2021}.

	For prokaryotic cells, the question of synchronization is especially important for   peritrichous bacteria equipped with multiple propulsion-inducing flagella, such as the model organism {\it Escherichia coli}~\cite{berg_book}: their helical flagellar filaments rotate passively under the actuation of molecular motors embedded in the cell wall and form coherent bundles behind the cell body, pushing the bacterium forward when it swims in a straight `run'~\cite{Turner2000}. The geometrical constraints imposed by the helical shape require the flagellar filaments to be in phase with each other to form smoothly rotating bundles~\cite{Macnab1977}. The level of synchronization between filaments is expected to influence the propulsive efficiency of the bundle, while intermittent loss of synchronization may contribute to the initiation of `tumble' events 
	%where the cells rapidly change directions
	~\cite{Darnton2007,Reigh2013}.

	Compared to eukaryotic flagella, the fundamental mechanisms of synchronization in rotating bacterial flagella are not  yet fully understood.
%Compared to eukaryotic flagella, the synchronization of rotating bacterial flagella has yet to be fully understood from a fundamental point of view.
 Computational studies have shown that some form of elastic compliance is necessary in addition to hydrodynamic interactions, since hydrodynamically-coupled helices rotating rigidly about a fixed axis do not synchronize~\cite{Kim2004b,Reichert2005}. Simulations have also revealed that the balance between bundling and synchronization times depends strongly on the initial separation between the filaments~\cite{Reigh2012} and that filaments may slip out of synchrony if driven by unequal torques~\cite{Reigh2013}.  Experimental studies on the synchronization of rotating bodies include systems of light-driven microrotors~\cite{DiLeonardo2012} and of macroscopic-scale rotating paddles~\cite{Qian2009}. 
	
\begin{figure}[t] 
		\includegraphics[width=\columnwidth]{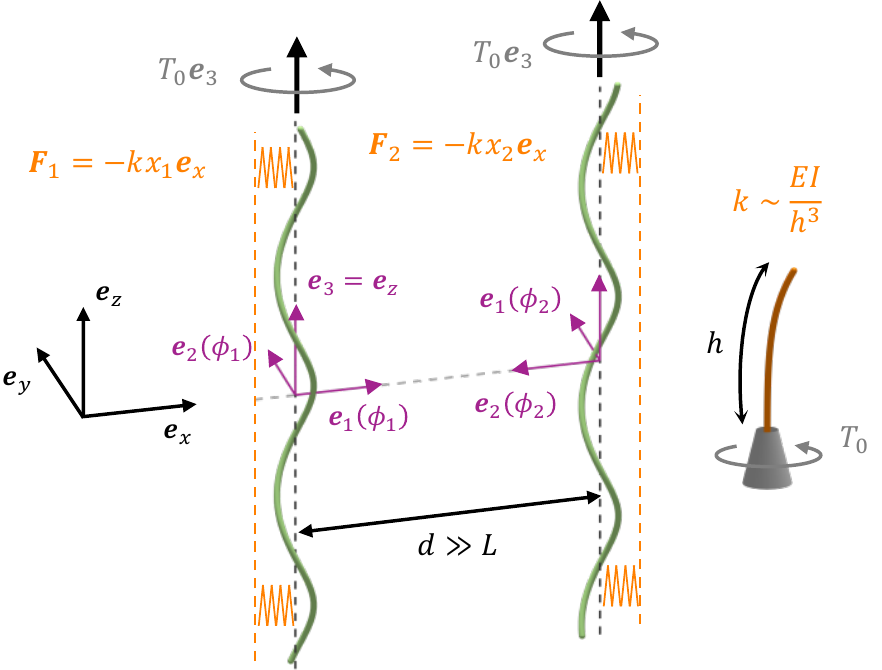}
		\caption{Reduced  model of interacting bacterial flagella. Two parallel helical filaments have phase angles $\phi_j$. A constant torque, $T_0\ez$, rotates each filament about its axis, while an elastic restoring force, $-kx_j\ex$, pulls the axis back to a reference position. The tethering points are separated along $x$ by a distance much greater than the length of the filaments, $d\gg L$.}
		\label{fig:setup}
	\end{figure}

\begin{figure}[t] 
		\includegraphics[width=\columnwidth]{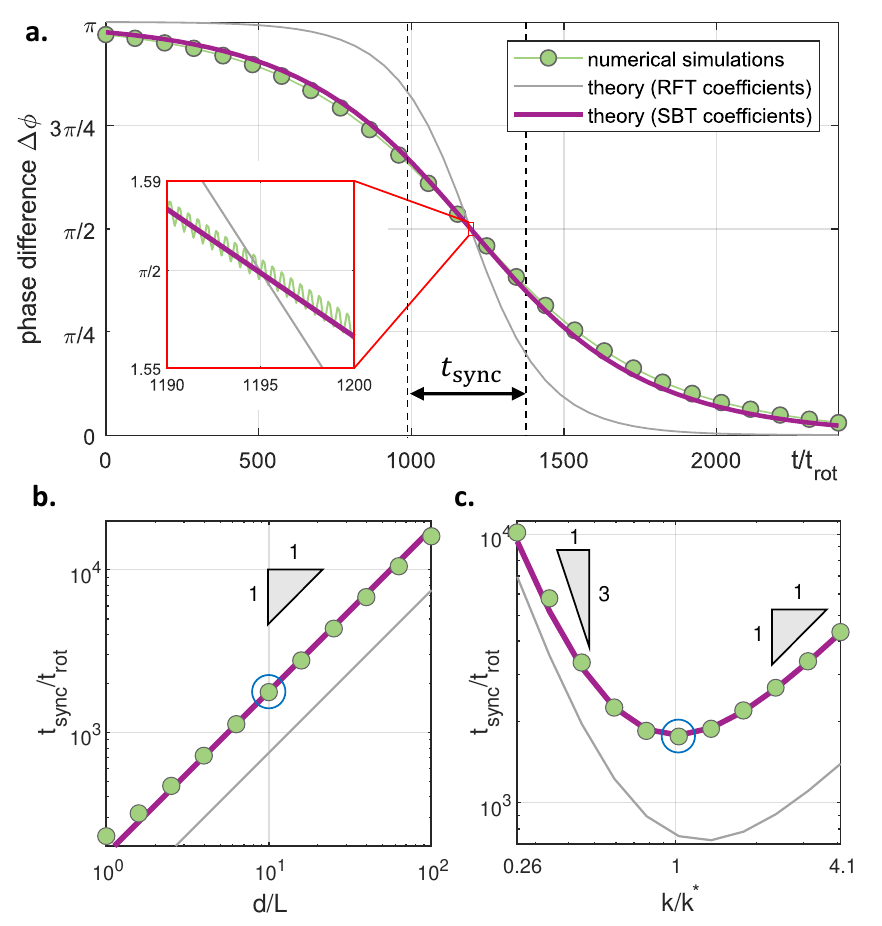}
		\caption{Interacting bacterial flagella generically synchronize in-phase via  an elastohydrodynamic mechanism. (a) Evolution of phase difference on the slow time scale of synchronization and (inset) on the fast time scale of rotation, {for inter-filament distance $d=2L$}. The time scale for synchronization is (b) proportional to the hydrodynamic coupling between the filaments
			and (c) minimized at intermediate elastic compliance. Data points circled in blue represent identical  input parameters (pitch angle $\psi = 0.446$ and filament thickness $\epsilon = 0.0038$ correspond to a ``normal'' flagellar filament~\cite{Turner2000}; {full list of parameters available in Supplementary Material~\cite{SM}}).}
		\label{fig:results}
	\end{figure}

\begin{figure}
		\includegraphics[width=\columnwidth]{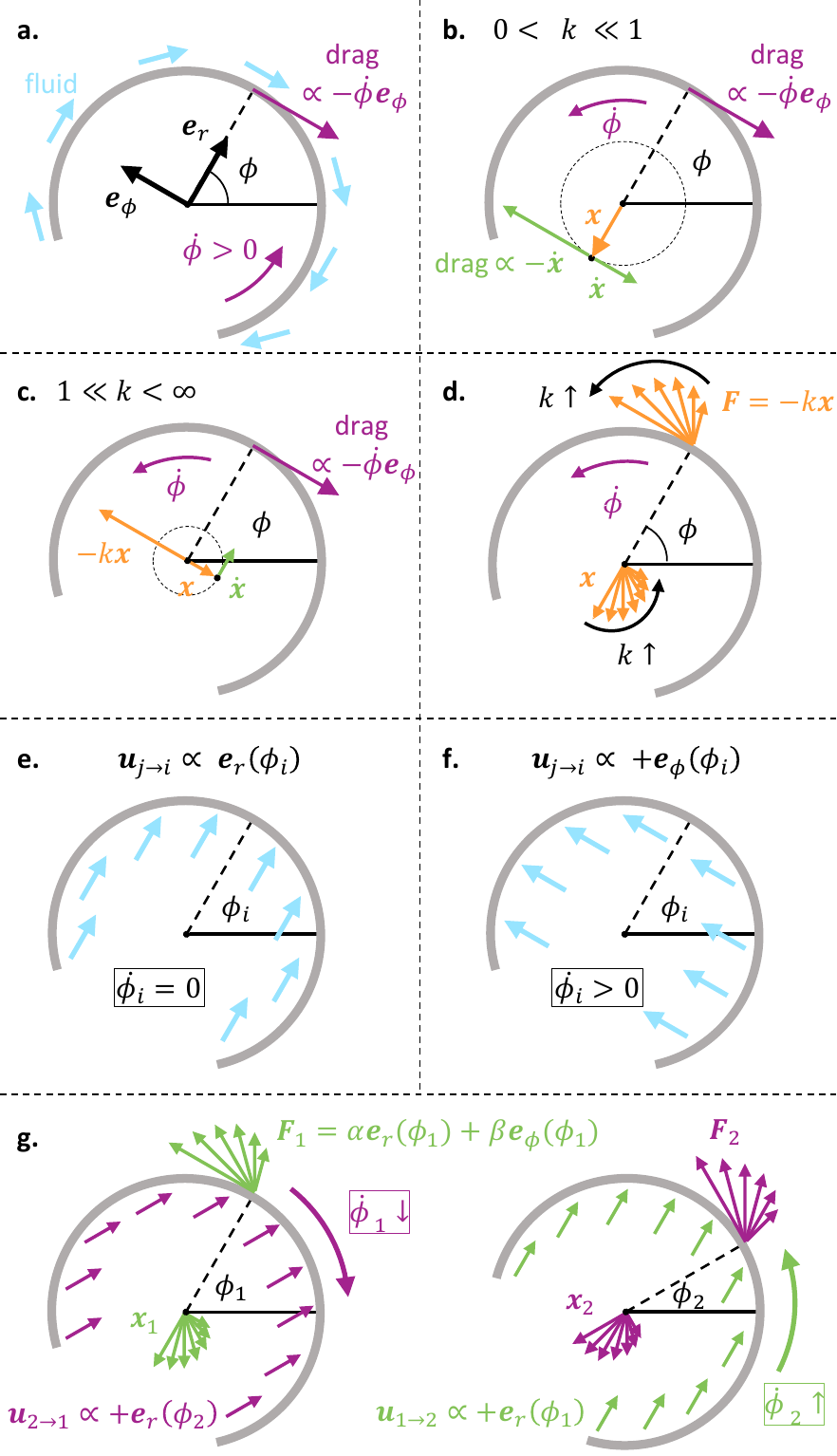}
		\caption{Physical mechanism for elastohydrodynamic synchronization. (a) The viscous drag on a rotating helix with phase angle $\phi$ is balanced out primarily by (b) the viscous drag due to translation ($k\ll 1$) or (c) by the elastic restoring force ($k\gg 1$). (d) For general values of $k$, the axis of the helix, $\xb$, lags behind the phase angle $\phi$ by an angle between $\pi/2$ and $\pi$. (e,f) Filament $i$ speeds up only if the flow induced by filament $j \neq i$ has a positive $\ephi(\phi_i)$ component.
			(g) Hydrodynamically-coupled helices synchronize in phase due to the flows that each filament $j$ produces along the positive $\er(\phi_j)$ direction and the resulting hydrodynamic stresses on filament $i\neq j$.}
		\label{fig:physmech}
	\end{figure}

	In this letter, we provide a microscopic physical model for the dynamics of compliant bacteria flagella
	% propose a reduced model of the bacterial flagellum
	and use the method of multiple scales to rigorously coarse-grain the equations of motion for {two flagella interacting in the far field} into an evolution equation for their mean phase difference. The model, illustrated in Fig.~\ref{fig:setup},
	preserves the salient features of the bacterial flagellum: the rotary motor operating near constant torque~\cite{Berry2008}, the left-handed helical shape of the semi-rigid flagellar filament~\cite{Hasegawa1998}, and the flexible connection between motor and filament via the flagellar hook~\cite{Sen2004}.
	In contrast to empirical models, our bottom-up approach includes all geometrical and dynamical details, thereby allowing us to derive the explicit dependence of the synchronization rate on the shape and dynamical properties of the bacterial flagellum. The theoretical predictions of our model are verified by numerical simulations (Fig.~\ref{fig:results}), {are shown to remain relevant beyond the far-field limit~\cite{SM}} and are elucidated by a physically intuitive explanation of the mechanism for synchronization, which  emerges from the elastohydrodynamic balance on individual filaments (Fig.~\ref{fig:physmech}). 
	Remarkably, we show that the rate of synchronization is maximised at intermediate values of elastic compliance, { with} significant implications for the biophysics of swimming bacteria. 
	
	To model the {far-field} interactions between bacterial flagella, we consider two identical and parallel rigid helices of length $L$, each rotating in a viscous fluid  (viscosity $\mu$) under a constant torque $T_0$. The reference positions of the axes are separated by a distance $|\db| = d \gg L$ along the $x$ direction (see Fig.~\ref{fig:setup}). We rescale lengths, forces and time such that $L=2$, $\mu=1$ and $T_0 = 1$ in dimensionless terms. On the {microscopic scale} of bacteria, the Reynolds number{s are} very small  
	so we operate within the framework of Stokes flow, { where} the relationship between the dynamics (forces $\Fb_j$ and torques $\Tb_j$) and kinematics (linear and angular velocities  $\Ub_j$ and  $\Omb_j$) of rigid bodies  is linear. For two flagellar filaments with phase angles $\phi_1$ and $\phi_2$, this means
	\begin{equation}
		\begin{pmatrix}
			\Fb_1 \\ \Tb_1
		\end{pmatrix} = \Sb(\db, \phi_1,\phi_2)\begin{pmatrix}
			\Ub_1 \\ \Omb_1
		\end{pmatrix} + \Cb(\db, \phi_1,\phi_2) 
		\begin{pmatrix}
			\Ub_2 \\ \Omb_2 
		\end{pmatrix},
		\label{eq:full-eqns-motions}
	\end{equation} 
	where the resistance matrices $\Sb$ and $\Cb$ provide the self-induced dynamics and the cross-interactions between filaments. In  previous work~\cite{PRF2021}, we demonstrate that $\Sb(\db, \phi_1,\phi_2) = \Sb_0(\phi_1) + \mathcal{O}(d^{-2})$ when $d\gg L = 2$, where $\Sb_0(\phi)$ is the resistance matrix for a filament in an infinite fluid, and is obtained by rotating $\Sb_0(0)=(\Ab, \Bb; \Bb^T, \Db)$ through an angle $\phi$ about the vertical axis.
	Meanwhile $\Cb(\db, \phi_1,\phi_2) = d^{-1}\Cb_1(\ex, \phi_1,\phi_2) + \mathcal{O}(d^{-2})$ and the first-order correction comes from the leading-order expansion of the Oseen tensor, $\Cb_1(\ex, \phi_1,\phi_2) = -\Sb_0(\phi_1)(\mathbf{I}+\ex\ex)\Sb_0(\phi_2)/(8\pi\mu)$. 
	%Implicitly, the first/last matrix in the product contains only the first three columns/rows of the full $\Sb_0$. 
	
	To model the elastic link between the flagellar filament and the rotary motor via  the flexible hook, as well as the elastic compliance of the semi-rigid filaments, 
	we allow the axis of each filament to move along $x$ { while} a linear elastic force (dimensionless strength $k$) { restores it to a} reference position (see Fig.~\ref{fig:setup}). The kinematics of helix $j$ is therefore described by two degrees of freedom:  lateral displacement, $x_j$, and   phase, $\phi_j$.  Projecting   Eq.~\eqref{eq:full-eqns-motions} onto the four degrees of freedom in our model, we obtain the reduced system up to $\mathcal{O}(d^{-1})$,
	\begin{equation}
		\begin{pmatrix}
			-kx_1 \\ T_0
		\end{pmatrix} = \tilde{\Sb}_0(\phi_1)\begin{pmatrix}
			\dot{x}_1 \\ \dot{\phi}_1
		\end{pmatrix} + d^{-1} \tilde{\Cb}_1(\ex, \phi_1,\phi_2)
		\begin{pmatrix}
			\dot{x}_2 \\ \dot{\phi}_2
		\end{pmatrix},
		\label{eq:resistance-matrix-simplified}
	\end{equation} 
	where tildes denote the appropriate subset of rows and columns from the matrices $\Sb_0$ and $\Cb_1$.

	We next exploit the separation of time scales that occurs at large inter-filament distance between fast rotation  and slow synchronization. We introduce a slow variable $\tau = d^{-1}t$ alongside the fast variable $t$ for rotation. The solution is then formally expanded as $\phi_j(t) =\phi_j^{(0)}(t,\tau) + d^{-1} \phi_j^{(1)}(t,\tau) + \mathcal{O}(d^{-2})$ with time derivative $\dot{\phi}_j(t) = \phi_{j,t}^{(0)} + d^{-1} (\phi_{j,\tau}^{(0)} + \phi_{j,t}^{(1)}) + \mathcal{O}(d^{-2})$, where subscripts $t$ and $\tau$ denote partial derivatives; a similar multiple scale expansion applies to lateral displacements, $x_j$. Following the method of multiple scales~\cite{Hinch1991}, we solve Eq.~\eqref{eq:resistance-matrix-simplified} at zero and first order in $d^{-1}$. At each order in $d^{-1}$ we further expand the solution in powers of a nested parameter $(\pi N)^{-1} \ll 1$, with $N$ being the number of helical turns (for a typical bacterial flagellar filament $(\pi N)^{-1} \approx 0.1$). This allows us to integrate the equations at each order analytically {(full calculations available in Supplementary Material~\cite{SM})}.
	
	At leading order in $d^{-1}$,the solution of Eq.~\eqref{eq:resistance-matrix-simplified} is that each filament rotates with fixed angular velocity $\Omega_0 = T_0 D_{33}^{-1}$ as $\phi_j^{(0)}(t,\tau) = \phih_j(\tau) + \Omega_0t$, where $\phih_j(\tau)$ is a constant of integration. Meanwhile, each filament axis oscillates as $x^{(0)}_j(t,\tau) \approx \rho(K)\cos(\phi^{(0)}_j-\xi(K))$ with amplitude $\rho(K) = B_{23}A_0^{-1}(K^2+1)^{-1/2}$ and phase lag $\xi(K) = \tan^{-1}(-K) \in (\pi/2,\pi)$ behind $\phi^{(0)}_j$, where $A_{ij}, B_{ij}, D_{ij}$ are the components of $\Sb_0(0)$ and $A_{0}=(A_{11}+A_{22})/2$.
	The dimensionless parameter $K=(T_0^{-1}D_{33})/(k^{-1}A_0) \equiv t_\mathrm{rot}/t_\mathrm{elast}$ encapsulates the dynamics of actuation (elastic compliance and driving torque) 
	and represents the ratio between the rotation time scale and the elastic relaxation time scale of the flagellum.

	At first order in $d^{-1}$, Eq.~\eqref{eq:resistance-matrix-simplified} can be reduced to a system for $\phi_1^{(0)}$ and $x_1^{(1)}$ only, by substituing the leading-order solution for $x^{(0)}_j$ and applying the solvability condition that the first-order correction $\phi_1^{(1)}$ remains bounded for large $t$. After some algebra, we deduce the slow evolution of the phase difference, $\Delta\phi = \phi_2^{(0)} - \phi_1^{(0)}$, to be  $\Delta\phi_\tau = -  kB_{23}(A_0D_{33})^{-1}(x_1^{(1)}\sin\phi_1^{(0)}-x_2^{(1)}\sin\phi_2^{(0)})$,
	where the first-order perturbation to sideways displacements	has the general solution $x_1^{(1)}(t,\tau) = \gamma \sin(\phi_1^{(0)}) + \delta \cos(\phi_1^{(0)}) + \zeta \sin(\phi_2^{(0)}) + \eta \cos(\phi_2^{(0)})$ once the initial transients have decayed exponentially. 
	
	By averaging { the evolution equation for $\Delta\phi$} over the fast time scale, we exactly recover the classical Adler equation \cite{Adler1946} for the mean phase difference $\langle\Delta\phi\rangle_\tau  = - kB_{23}(A_0D_{33})^{-1} \eta\sin(\Delta\phi) \equiv -\tau_\mathrm{sync}^{-1}\sin(\Delta\phi).$ This leads generically to in-phase synchronization as $\langle\Delta\phi\rangle = 2\tan^{-1}\left(\tan(\Delta\phi_0/2)\exp(-\tau/\tau_{\mathrm{sync}})\right)$ 
	(Fig.~\ref{fig:results}a). Importantly, by identifying the coefficient $\eta$ in the solution for $x_1^{(1)}$, we can predict the time scale for synchronization $t_{\mathrm{sync}} = d\tau_{\mathrm{sync}} = 2\pi\mu d D_{33}^2(B_{23}^2T_0)^{-1}(K^2 + 1)^2K^{-3}$, which depends explicitly on the actuation dynamics through $K$ and on the helical shape through the individual viscous resistance coefficients $A_0,B_{23},D_{33}$. 
	Notably, {$t_{\mathrm{sync}} \propto (K^2 + 1)^2K^{-3}$ is minimized} at an optimum value, $K^* = \sqrt{3}$, independent of distance.

	To validate our multiple-scale approach, we compute the instantaneous hydrodynamic forces on the two filaments using Johnson's slender-body theory (SBT)~\cite{Johnson1980} and then time-step  using  Runge-Kutta RK4. We include the full hydrodynamic interactions between the filaments by solving the integral equation $8\pi\mu\ub_1(s) = \mathcal{L}[\fb_1(s)] + \mathcal{K}[\fb_1(s')] + \mathcal{J}[\fb_2(s'),\db]$, where $\ub_j$ and $\fb_j$ are the velocity and force density along filament $j$, $\mathcal{L}$ and $\mathcal{K}$ represent local and non-local effects on the same filament, and the integral operator $\mathcal{J}$ contains the Stokeslet and source dipole flows generated by the other filament~\cite{Tornberg2004}. 
	The integral equation is solved numerically using a Galerkin method with Legendre polynomials~\cite{PRF2021}. {Further details of the computational method are available in the Supplementary Material~\cite{SM}.}
	The final { expression for $t_{\mathrm{sync}}$}, with $A_0,B_{23},D_{33}$ evaluated either analytically via resistive-force theory (RFT)~\cite{Hancock1953,Gray1955,Lighthill1976} or computationally via SBT ~\cite{Johnson1980}, is compared against full numerical simulations in Fig.~\ref{fig:results}. 
	Our multiple-scale theory with SBT coefficients is in perfect agreement with simulations, while RFT captures all qualitative features of synchronization.
	
	{ Our   model of interacting bacterial flagella  was  coarse-grained using  multiple scales and shown to reduce to the Adler equation, leading to in-phase locking. }	What is the physical mechanism    responsible for this  flagellar synchronization, and why is synchronization    fastest at intermediate values of elastic compliance? 

	To convey   physical intuition, we focus on the details of   hydrodynamic forces { and} allow the filaments to move in a 2D elastic trap (the 1D/2D trap were confirmed to be qualitatively identical via numerical simulations). Since both rotation and lateral displacement occur in the plane perpendicular to the filament axis, it {suffices} to consider the flows and forces acting on the horizontal projection of the helical centreline. When projected, each helical flagellar filament maps onto a circle. { If the helix has a non-integer number of turns, we have a surplus of filament on one side of the circle, so the filament generically reduces} to the arc of a circle, or a ``horseshoe'' (Fig.~\ref{fig:physmech}).

	We first describe, in Fig.~\ref{fig:physmech}a-d, the intrinsic dynamics of an elastically-tethered and rotating horseshoe in the absence of hydrodynamic interactions. A rotating horseshoe with phase angle $\phi$  defined as in Fig.~\ref{fig:physmech}a experiences a net viscous drag in the negative $\ephi(\phi)$ direction due to a one-sided surplus in force (Fig.~\ref{fig:physmech}a). For weak elastic stiffness ($k\ll 1$, Fig.~\ref{fig:physmech}b), this viscous drag due to rotation is balanced out primarily by the viscous drag due to translation, and thus we have  $\dot{\xb} \propto -\ephi(\phi)$. In contrast,  for strong elastic stiffness ($k\gg 1$, Fig.~\ref{fig:physmech}c), the viscous drag from rotation is balanced primarily by the elastic restoring force, and therefore we have  $\xb \propto -\ephi(\phi)$.  In the intermediate regime ($k=O(1)$, Fig.~\ref{fig:physmech}d), the centre of the projected filament oscillates on circular orbits lagging behind the phase $\phi$ by an angle between $\pi/2$ and $\pi$. 
	Importantly, the force exerted by the filament on the fluid, $ \Fb = -k \xb=\alpha\er(\phi) + \beta\ephi(\phi)$, always has a positive $\er$ component.

To reveal the evolution of the phase difference for interacting helices, 
	we consider in  Fig.~\ref{fig:physmech}e-g how  horseshoe $i$  responds to the flows induced by   horseshoe $j\neq i$. If the external flow created by filament $j$ near  filament $i$  is parallel to $\er(\phi_i)$, then the hydrodynamic forces are balanced by symmetry and the external flow does not lead to additional rotation (Fig.~\ref{fig:physmech}e). If, however, the flow induced by  filament $j$ has a positive $\ephi(\phi_i)$ component, then  filament $i$ experiences a hydrodynamic torque and speeds up
	(Fig.~\ref{fig:physmech}f). In the far field, the flow induced by   filament $j$ at the position of   filament $i$ is a uniform flow $\Fb_j\cdot(\Ib+\ex\ex)/8\pi\mu d$ (the leading order expansion of a Stokeslet flow).  The $\ephi(\phi_j)$ component of $\Fb_j$ leads to a symmetric term $ \beta\ephi(\phi_i)\cdot(\Ib+\ex\ex)\cdot\ephi(\phi_j)$ on both filaments that does not modify the phase difference. In contrast, the flow induced by the positive $\er(\phi_j)$ component of $\Fb_j$  does lead to synchronization: as sketched  in Fig.~\ref{fig:physmech}g, these flow components have the effect of slowing down the filament that is ahead and speeding up the one that is behind. This elastohydrodynamic balance is the  physical mechanism responsible for the synchronization of bacterial filaments.  
	
	The phase dynamics illustrated theoretically and numerically in Fig.~\ref{fig:results}b-c are fully explained by the physical mechanism outlined above: (i) the rate of synchronization decays linearly with $d$, a signature of Stokeslet flows, and 
	(ii) the rate of synchronization is proportional to the radial component of the force, $\alpha=\Fb_j\cdot\er(\phi_j)$, which is largest for intermediate values of $k$ since $\alpha\ll 1$ when either $k\ll 1$ or $k\gg 1$, because $\Fb_j\approx 0$ or $\Fb_j \propto \ephi(\phi_j)$ respectively (Fig.~\ref{fig:physmech}d).

	Previous biophysical models of ciliary synchronization  assumed  $K = t_\mathrm{rot}/t_\mathrm{elast}$ to be very large~\cite{Niedermayer2008,Qian2009,Tanasijevic2021}.
	While this is suitable for eukaryotic flagella, the dynamical properties of bacterial flagella are qualitatively different. Crucially, the bacterial flagellum has two components with bending stiffness separated by four orders of magnitude: $EI_{hook} \approx 1.6\times10^{-4}$ pN \um$^2$~\cite{Sen2004} and $EI_{filament} \approx 3.5$ pN \um$^2$~\cite{Darnton2007b}. 
	On dimensional grounds, we have $k\sim EI/h^3$, where $EI$ and $h$ are the bending stiffness and length of the deforming component (hook or filament). We estimate that $K_{hook}\approx 0.05$, which is significantly below the optimum $K^*=\sqrt{3}$ and gives $t_{\mathrm{sync}}/t_{\mathrm{opt}}=\mathcal{O}(10^{3})$ due to the rapid increase of $t_{\mathrm{sync}}\sim K^{-3}$ for small $K$. However, the finite size of the hook ($h = 59$ nm~\cite{Sen2004}) likely requires that the proximal end of the flagellar filament deforms as well. For a length scale of deformation { around} $h = 0.1-1$~\um, we estimate that $K_{filament} = 0.2 - 200$, a range which straddles the optimum. Surprisingly, this suggests that the comparatively smaller elastic compliance of the semi-rigid flagellar filament \cite{Darnton2007b}, together with the short length \cite{Sporing2018} and the dynamic stiffening of the flagellar hook \cite{nord2021dynamic} could play a crucial role in the synchronization and stability of bacterial flagellar bundles.

	The physical mechanism revealed by our reduced model of the bacterial flagellum is distinct from previously proposed eukaryotic mechanisms based on orbital compliance~\cite{Niedermayer2008} and phase-dependent forcing~\cite{Uchida2012}. It is similar  to the axial-compliance mechanism in Ref.~\cite{Qian2009}, but the resulting dynamics of synchronization are qualitatively different due to the specific hydrodynamic resistance of the helical filament. Since the helical amplitude of a bacterial flagellar filament  is  much smaller than its length, its resistance to rotation, $D_{33}$, is much smaller than its resistance to translation, $A_0$. This leads to  intrinsic kinematics of an individual filament (Fig.~\ref{fig:physmech}b-d) that cannot be captured by spherical-bead models.  
	
	{ One limitation of our analytical theory is that each filament moves with only two degrees of freedom. Computations for filaments with all six degrees of freedom~\cite{SM} reveal that the physical mechanism for synchronization is robust against deviations in the axes of the filaments, and that our result {for $t_{\mathrm{sync}}$} can be modified to take into account a finite angle of inclination between the two filaments (as observed during flagellar bundling). Crucially, the optimum elastic compliance for synchronization is not affected by the inclination.
		
	Another constraint that can be relaxed, even analytically, is the assumption that the filaments are identical~\cite{SM}. The Adler equation can be modified to include a small mismatch, $\Delta\Omega_0$, in the intrinsic rotation rates of the two filaments, giving $\langle\Delta\phi\rangle_\tau  = \Delta\Omega_0 -\tau_\mathrm{sync}^{-1}\sin(\Delta\phi)$. The filaments then phase-lock to $\Delta\phi_\infty = \sin^{-1}\left(\Delta\Omega_0\tau_\mathrm{sync}\right)$, up to a critical point where the mismatch cannot be compensated by hydrodynamic interactions. This  prediction agrees with numerical simulations for two filaments with either mismatched driving torques or different filament geometries.}

	{To make analytical progress, it is often necessary to assume that   hydrodynamically-coupled bodies are far apart~\cite{Niedermayer2008,Qian2009,Uchida2011,Brumley2014}.}
	{Despite such constraints,  far-field theories are crucial for understanding the underlying principles of synchronization. Here, the theory allowed us to} provide a detailed  physical mechanism based on an elastohydrodynamic balance at the level of individual filaments, { which remains valid beyond the far-field limit~\cite{SM}}. The current theory and simulations highlight the essential role played by the elasticity of both the hook and the flagellar filament in the synchronization of bacterial flagella. {Numerical simulations beyond the far-field limit~\cite{SM}  suggest that the far-field optimum value of $K$ could distinguish between different near-field trends for $t_{\mathrm{sync}}$. In turn, this could guide future studies into the synchronisation of bacterial flagella separated by distances smaller than the filament length, e.g.~during flagellar bundling.}
	Our results will enable a comparative analysis of synchronization for the different polymorphic shapes of bacterial flagellar filaments~\cite{Hasegawa1998}, while our methodology can also be used to  investigate the impact of the  torque-speed relationship of the rotary motor on   synchronization~\cite{Berry2008}. {Finally, the  noisy Adler equation  has been proposed to model the fluctuating dynamics of beating filaments~\cite{Chakrabarti2019,Goldstein2009}  
 so	  adding noise  to the system} studied here could enable a better understanding of flagellar unbundling~\cite{Turner2000}.
	
	\begin{acknowledgments}
		 We gratefully acknowledge funding from the University of Cambridge (George and Lillian Schiff Studentship supporting M.T.C.), from Collège de France (A.T.E.R. contract supporting M.T.C.), and from the European Research Council through the European Union’s Horizon 2020 research and innovation programme (under grant agreement No. 682754 to E.L.).
	\end{acknowledgments}

	%\bibliography{mybibliography}
	

\end{document}

\title{Supplementary Material for \\ ``Elastohydrodynamic synchronization of rotating bacterial flagella"}

\author{Maria T\u{a}tulea-Codrean}
%\email[]{m.tatulea-codrean@damtp.cam.ac.uk}
\affiliation{Department of Applied Mathematics and Theoretical Physics, University of Cambridge, Cambridge CB3 0WA, United Kingdom}
\affiliation{Coll\`{e}ge de France, 11 place Marcelin Berthelot, 75005 Paris, France}
\author{Eric Lauga}
\email[]{e.lauga@damtp.cam.ac.uk}
\affiliation{Department of Applied Mathematics and Theoretical Physics, University of Cambridge, Cambridge CB3 0WA, United Kingdom}

\def\v{\vspace{2cm}}
\date{\today}

\maketitle

In Section \ref{sec:theory} of this Supplementary Material, we provide the background calculations that take us from the initial governing equations, Eq.~(2) in the main manuscript, to the final result concerning the time scale for synchronization. {In Section \ref{sec:comp-method}, we describe the computational method used for all the numerical simulations in the main manuscript and the supplementary material}. Next, in Section \ref{sec:simulations}, we present evidence from extensive numerical simulations in which we relax some limiting assumptions of our analytical theory, in order to demonstrate the usefulness of our analytical results for more general synchronization conditions. Finally, in Section \ref{sec:parameters}, we provide a systematic list of the simulation parameters for each figure in the main manuscript and the supplementary material.

\tableofcontents

\vspace{\baselineskip}

\section{Analytical model of synchronization}
\label{sec:theory}

\subsection{Governing equations}

Working within the setup introduced in the main manuscript, the equations of motion for two elastically-tethered and hydrodynamically-coupled rotating helices with phase angles $\phi_1$ and $\phi_2$ are
\begin{equation}
\begin{pmatrix}
-kx_1 \\ T_0
\end{pmatrix} = \tilde{\Sb}_0(\phi_1)\begin{pmatrix}
\dot{x}_1 \\ \dot{\phi}_1
\end{pmatrix} + d^{-1} \tilde{\Cb}_1(\ex, \phi_1,\phi_2)
\begin{pmatrix}
\dot{x}_2 \\ \dot{\phi}_2
\end{pmatrix},
\label{eq:resistance-matrix-simplified}
\end{equation}
where the left-hand side of the equation represents the prescribed dynamics on the first filament (an elastic restoring force, of strength $k$, and a constant driving torque, $T_0$) and the right-hand side represents the viscous force and torque exerted by the filament on the fluid due to its own motion, through the self-induced resistance matrix $\tilde{\Sb}_0(\phi_1)$, and due to the motion of the second filament, through the cross-interaction matrix, $\tilde{\Cb}_1(\ex, \phi_1,\phi_2)$. The equations of motion for the second filament are immediately obtained from Eq.~\eqref{eq:resistance-matrix-simplified} by swapping the indices $1\leftrightarrow 2$ and sending $\ex \mapsto - \ex$.

The tildes in Eq.~\eqref{eq:resistance-matrix-simplified} indicate that an appropriate subset of rows and columns has been taken from the full resistance matrices, $\Sb_0$ and $\Cb_1$, such that
\begin{equation}
    \tilde{\mathbf{S}}_0(\phi_1) = \begin{pmatrix}
S^{(0)}_{11}(\phi_1) & S^{(0)}_{16}(\phi_1) \\
S^{(0)}_{61}(\phi_1) & S^{(0)}_{66} 
\end{pmatrix}, \quad \tilde{\mathbf{C}}_1(\ex,\phi_1,\phi_2) = \begin{pmatrix}
C^{(1)}_{11}(\phi_1,\phi_2) & C^{(1)}_{16}(\phi_1,\phi_2) \\
C^{(1)}_{61}(\phi_1,\phi_2) & C^{(1)}_{66}(\phi_1,\phi_2) 
\end{pmatrix}. 
\end{equation}
The self-induced resistance matrix for a filament with arbitrary phase angle, $\phi$, is obtained by rotating the resistance matrix for zero phase angle, $\Sb_0(0)=(\Ab, \Bb; \Bb^T, \Db)$, by an angle $\phi$ about the vertical 
\begin{equation}
\mathbf{S}_0(\phi) = \begin{pmatrix}
\mathbf{Q}\mathbf{A}\mathbf{Q}^T & \mathbf{Q}\textbf{B}\mathbf{Q}^T \\ \mathbf{Q}\mathbf{B}^T\mathbf{Q}^T & \mathbf{Q}\mathbf{D}\mathbf{Q}^T
\end{pmatrix} \quad \mathbf{Q}(\phi) = 
\begin{pmatrix}
\cos\phi & -\sin\phi & 0 \\
\sin\phi & \cos\phi & 0 \\
0 & 0 & 1
\end{pmatrix}.
\label{eq:changeofbasis}
\end{equation}
The coefficients $A_{ij}, B_{ij}, D_{ij}$ of the resistance matrix $\Sb_0(0)$ were calculated via resistive-force theory (RFT) in previous work \cite{PRF2021}.

It was also shown, in previous work \cite{PRF2021}, that the cross-interaction resistance matrix is given at order $\mathcal{O}(d^{-1})$ by the leading-order expansion of the Oseen tensor, $\Cb_1(\pm\ex, \phi_1,\phi_2) = -\Sb_0(\phi_1)(\mathbf{I}+\ex\ex)\Sb_0(\phi_2)/(8\pi\mu)$. Expanding this out into components, we have
\begin{equation}
C_{ij}^{(1)} (\phi_1,\phi_2) = -\frac{2S^{(0)}_{i1}(\phi_1)S^{(0)}_{1j}(\phi_2)+S^{(0)}_{i2}(\phi_1)S^{(0)}_{2j}(\phi_2)+S^{(0)}_{i3}(\phi_1)S^{(0)}_{3j}(\phi_2)}{8\pi \mu},
\label{eq:cross-definition}
\end{equation}
where the indices $i, j \in \{1,2,\dots,6\}$. 

\subsection{Multiple scales expansion}

We solve Eq.~\eqref{eq:resistance-matrix-simplified} using the method of multiple scales \cite{Hinch1991}, whereby we introduce a slow time variable for synchronization, $\tau = d^{-1}t$, alongside the fast time variable for rotation, $t$. We treat the two time variables formally as independent of each other, and we express the phase angles and lateral displacements of the two helices as series expansions,
\begin{eqnarray}
x_j(t) &=& x_j^{(0)}(t,\tau) + d^{-1} x_j^{(1)}(t,\tau) + \mathcal{O}(d^{-2}), \label{eq:multiple-scale-expansion-x}\\
\phi_j(t) &=& \phi_j^{(0)}(t,\tau) + d^{-1} \phi_j^{(1)}(t,\tau) + \mathcal{O}(d^{-2}), \label{eq:multiple-scale-expansion-phi}
\end{eqnarray}
with time derivatives
\begin{eqnarray}
\dot{x}_j(t) &=& x_{j,t}^{(0)} + d^{-1} (x_{j,\tau}^{(0)} + x_{j,t}^{(1)}) + \mathcal{O}(d^{-2}), \label{eq:multiple-scale-expansion-x-derivative}\\
\dot{\phi}_j(t) &=& \phi_{j,t}^{(0)} + d^{-1} (\phi_{j,\tau}^{(0)} + \phi_{j,t}^{(1)}) + \mathcal{O}(d^{-2}). \label{eq:multiple-scale-expansion-phi-derivative}
\end{eqnarray}
where subscripts $t$ and $\tau$ denote partial derivatives.

\subsection{Derivation and solution of leading-order equations}
\label{sec:leadingorderkinematics}
Using the multiple-scales expansion from Eqs.~\eqref{eq:multiple-scale-expansion-x}-\eqref{eq:multiple-scale-expansion-phi-derivative}, the leading-order terms in the governing equations, Eq.~\eqref{eq:resistance-matrix-simplified}, are  
\begin{equation}
\begin{pmatrix}
x_{j,t}^{(0)} \\ \phi_{j,t}^{(0)}
\end{pmatrix} = \tilde{\mathbf{S}}_0^{-1}(\phi_j^{(0)}) \begin{pmatrix}
-kx_j^{(0)} \\ T_0
\end{pmatrix},
\label{eq:multiplescales-order0}
\end{equation}
We compute the inverse of the resistance matrix as a series expansion in the nested asymptotic parameter $(\pi N)^{-1} \ll 1$ (recall that for a typical bacterial flagellar filament $(\pi N)^{-1} \approx 0.1$), where $N$ is the number of helical turns of the helix. Hence, we have
\begin{eqnarray}
\tilde{\mathbf{S}}_0^{-1}(\phi) &=& \hat{\mathbf{S}}_2(\phi) + \hat{\mathbf{S}}_0(\phi) + \mathcal{O}\left((\pi N)^{-1}\right). \label{eq:Sinv-series}
\end{eqnarray}
This is possible due to a separation of scales between the coefficients of $\tilde{\mathbf{S}}_0$, obtained from Eq.~\eqref{eq:changeofbasis} as
\begin{eqnarray}
S^{(0)}_{11}(\phi) &=& A_0 + \Delta A \cos(2\phi), \\
S^{(0)}_{16}(\phi) &=& -B_{23}\sin(\phi) = S^{(0)}_{61}(\phi), \\
S^{(0)}_{66}(\phi) &=& D_{33},
\end{eqnarray}
where $A_{0}=(A_{11}+A_{22})/2$ and $\Delta A=(A_{11}-A_{22})/2$. The scaling of these terms with respect to the dimensionless parameter $(\pi N)^{-1} \ll 1$ can be inferred from RFT calculations \cite{PRF2021}, and we have
\begin{eqnarray}
 A_0 &\sim& \mathcal{O}(1), \\
 \Delta A &\sim& \mathcal{O}((\pi N)^{-1}),  \\
 B_{23}, D_{33} &\sim& \mathcal{O}((\pi N)^{-2}). 
\end{eqnarray}
Therefore, we obtain the first two terms in the series expansion of $\tilde{\mathbf{S}}_0^{-1}$ from Eq.~\eqref{eq:Sinv-series},
\begin{eqnarray}
\hat{\mathbf{S}}_2(\phi) &=& \begin{pmatrix}
~0 & 0 \\ ~0 & D_{33}^{-1}
\end{pmatrix} \sim (\pi N)^{2}, \label{eq:matrix-Shat2}
\\
\hat{\mathbf{S}}_0(\phi) &=& \begin{pmatrix}
A_0^{-1} & A_0^{-1}B_{23}D_{33}^{-1}\sin(\phi) \\ A_0^{-1}B_{23}D_{33}^{-1}\sin(\phi)  &  A_0^{-1}\left(B_{23}D_{33}^{-1}\sin(\phi)\right)^2 
\end{pmatrix} \sim (\pi N)^{0}. \label{eq:matrix-Shat0}
\end{eqnarray}

By substituting the above expansion of $\tilde{\mathbf{S}}_0^{-1}$ into Eq.~\eqref{eq:multiplescales-order0} we deduce that, to leading order, the rate of change of the phase difference is 
\begin{equation}
    \frac{\partial \phi_j^{(0)}}{\partial t} = T_0 D_{33}^{-1} + \mathcal{O}((\pi N)^0), \label{eq:phi-intrinsic}
\end{equation}
with solution
\begin{equation}
    \phi_j^{(0)}(t,\tau)  = \hat{\phi}_j(\tau) + T_0 D_{33}^{-1}t.
    \label{eq:phi-leadingorder}
\end{equation}
Likewise, the leading-order equation for the lateral displacement is
\begin{equation}
\frac{\partial x_j^{(0)}}{\partial t} = -\frac{k}{A_0}x_j^{(0)} + \frac{B_{23} T_0}{A_0 D_{33}}\sin(\phi_j^{(0)}) + \mathcal{O}((\pi N)^{-1}). \label{eq:x-intrinsic}
\end{equation}
Eq.~\eqref{eq:x-intrinsic} can be integrated by parts after multiplying both sides of the equation by an appropriate integrating factor, $\exp(kt/A_0)$, to find that
\begin{equation}
x_j^{(0)} \approx \frac{B_{23}}{A_0} \times \frac{K\sin(\phi_j^{(0)}) - \cos(\phi_j^{(0)})}{K^2 + 1}, \quad K = \frac{kD_{33}}{A_0 T_0}. \label{eq:defn-K}
\end{equation}
This result can be rewritten in a more compact form as
\begin{equation}
    x_j^{(0)} \approx \rho(K)\cos(\phi_j^{(0)}-\xi(K)), \label{eq:xlagsbehindphi}
\end{equation}
where the amplitude and phase lag of the lateral displacement are
\begin{eqnarray}
\rho(K) = \frac{B_{23}}{A_0\sqrt{K^2+1}}, \qquad \xi(K) = \tan^{-1}(-K) \in \left(\frac{\pi}{2},\pi\right). \label{eq:amplitude_and_lag}
\end{eqnarray}
The fact that the phase lag falls in the interval $\xi \in \left(\frac{\pi}{2},\pi\right)$ is important for synchronization (see the discussion of the physical mechanism in the main manuscript).

\subsection{Derivation of first-order equations}

Using the multiple-scales expansion from Eqs.~\eqref{eq:multiple-scale-expansion-x}-\eqref{eq:multiple-scale-expansion-phi-derivative}, the first-order terms in the governing equations, Eq.~\eqref{eq:resistance-matrix-simplified}, are  
\begin{multline}
\begin{pmatrix}
x_{1,\tau}^{(0)} + x_{1,t}^{(1)} \\ \phi_{1,\tau}^{(0)} + \phi_{1,t}^{(1)}
\end{pmatrix} = \tilde{\mathbf{S}}_0^{-1}(\phi_1^{(0)})\begin{pmatrix} -kx_1^{(1)} \\ 0
\end{pmatrix} + \phi_1^{(1)}\frac{\partial \tilde{\mathbf{S}}_0^{-1}}{\partial \phi}(\phi_1^{(0)})\begin{pmatrix}
-k x_1^{(0)} \\ T_0
\end{pmatrix} \\
-\tilde{\mathbf{S}}_0^{-1}(\phi_1^{(0)})\tilde{\mathbf{C}}_1(\ex,\phi_1^{(0)},\phi_2^{(0)})\tilde{\mathbf{S}}_0^{-1}(\phi_2^{(0)}) \begin{pmatrix}
-kx_2^{(0)} \\ T_0
\end{pmatrix}.
\label{eq:multiplescales-order1}
\end{multline}
We eliminate $x_{1}^{(0)}$ using the solution from Eq.~\eqref{eq:xlagsbehindphi} to find that
\begin{eqnarray}
\begin{pmatrix}
~1~ & -\rho\sin(\phi^{(0)}_1-\xi) \\ ~0~ & 1
\end{pmatrix}
\begin{pmatrix}
x_{1,t}^{(1)} \\ \phi_{1,\tau}^{(0)}
\end{pmatrix} 
+ 
\begin{pmatrix}
0 \\ \phi_{1,t}^{(1)}
\end{pmatrix} 
&=& 
\tilde{\mathbf{S}}_0^{-1}(\phi_1^{(0)})\begin{pmatrix} -kx_1^{(1)} \\ 0 \end{pmatrix} \nonumber
\\ &+& \phi_1^{(1)}\frac{\partial \tilde{\mathbf{S}}_0^{-1}}{\partial \phi}(\phi_1^{(0)})
\begin{pmatrix}
-k \rho\cos(\phi^{(0)}_1-\xi) \\ T_0
\end{pmatrix} \nonumber
\\
&-& \tilde{\mathbf{S}}_0^{-1}(\phi_1^{(0)})\tilde{\mathbf{C}}_1(\ex,\phi_1^{(0)},\phi_2^{(0)})\tilde{\mathbf{S}}_0^{-1}(\phi_2^{(0)}) 
\begin{pmatrix}
-k\rho\cos(\phi^{(0)}_2-\xi) \\ T_0
\end{pmatrix}. \label{app:eq:first-order-dynsys}
\end{eqnarray}
However, the above system of equations is under-determined since there are three unknowns, $\phi_1^{(0)}, x_1^{(1)}, \phi_1^{(1)}$, and only two equations. The standard way forward in the multiple-scales method is to make use of an additional constraint called a ``solvability condition'' in order to close the system \cite{Hinch1991}.

\subsection{Derivation of solvability condition}

In order for the multiple-scales expansion from Eq.~\eqref{eq:multiple-scale-expansion-phi} to be valid for all times, the first-order correction, $\phi_1^{(1)}$, must remain bounded for large times $t$. We gather all terms proportional to $\phi_1^{(1)}$ from Eq.~\eqref{app:eq:first-order-dynsys} to the left-hand side,
\begin{equation}
\begin{pmatrix}
0 \\ \phi_{1,t}^{(1)}
\end{pmatrix}
- \phi_1^{(1)}\frac{\partial \tilde{\mathbf{S}}_0^{-1}}{\partial \phi}(\phi_1^{(0)})
\begin{pmatrix}
-k \rho\cos(\phi^{(0)}_1-\xi) \\ T_0
\end{pmatrix}
= \begin{pmatrix}
f(t,\tau) \\ g(t,\tau)
\end{pmatrix},\label{app:eq:phi1}
\end{equation}
leaving us with two unknown functions $f(t,\tau)$ and $g(t,\tau)$ on the right-hand side, which depend on $\phi_1^{(0)}$ and $ x_1^{(1)}$.

From Eqs.~\eqref{eq:Sinv-series}, \eqref{eq:matrix-Shat2} and \eqref{eq:matrix-Shat0} we deduce that, to leading order,
\begin{equation}
    \frac{\partial \tilde{\mathbf{S}}_0^{-1}}{\partial \phi} = \frac{B_{23}\cos(\phi)}{A_0 D_{33}}\begin{pmatrix}
0 & 1 \\ 1 &  2B_{23}D_{33}^{-1}\sin(\phi)
\end{pmatrix}.
\end{equation}
Substituting this into Eq.~\eqref{app:eq:phi1}, we get that
\begin{eqnarray}
f(t,\tau) &=& -\frac{T_0B_{23}}{A_0 D_{33}}\cos(\phi_1^{(0)})\phi_1^{(1)}, \label{app:eq:f}\\
g(t,\tau) &=& \frac{\partial \phi_1^{(1)}}{\partial t} - \phi_1^{(1)}\frac{B_{23}\cos(\phi_1^{(0)})}{A_0 D_{33}} \left(-k \rho\cos(\phi^{(0)}_1-\xi) + \frac{2T_0 B_{23}}{D_{33}}\sin(\phi_1^{(0)})\right). \label{app:eq:g}
\end{eqnarray}
Suppose that the function $f(t,\tau)$ is non-zero. Then, from Eq.~\eqref{app:eq:f} it follows that $\phi_1^{(1)}$ is non-zero as well, so $\phi_1^{(1)}$ is a non-trivial solution of Eq.~\eqref{app:eq:g}. Regardless of the function $g(t,\tau)$, this solution involves the complementary function $\phi_{\mathrm{CF}}$ which solves the homogeneous equation
\begin{equation}
\frac{\partial \phi_{\mathrm{CF}}}{\partial t} - \frac{B_{23}\cos(\phi_1^{(0)})}{A_0 D_{33}} \left(-k \rho\cos(\phi^{(0)}_1-\xi) + \frac{2T_0 B_{23}}{D_{33}}\sin(\phi_1^{(0)})\right) \phi_{\mathrm{CF}} = 0.
\end{equation}
This equation is separable, so we integrate it as
\begin{equation}
    \int \frac{\mathrm{d}\phi_{\mathrm{CF}}}{\phi_{\mathrm{CF}}} = \frac{B_{23}}{A_0 D_{33}} \int \cos(\phi_1^{(0)})\left(-k \rho\cos(\phi^{(0)}_1-\xi) + \frac{2T_0 B_{23}}{D_{33}}\sin(\phi_1^{(0)})\right) \mathrm{d}t.
\end{equation}
The integral on the right-hand side is evaluated using the fact that $\phi_{1,t}^{(0)}=T_0D_{33}^{-1}$ at leading order (so $\mathrm{d}t \approx T_0^{-1}D_{33} \mathrm{d}\phi_{1}^{(0)}$). Then, using the definitions of $\rho$ and $\xi$ from Eq.~\eqref{eq:amplitude_and_lag} we deduce that the complementary function grows exponentially over the time,
\begin{equation}
    \phi_{\mathrm{CF}} \propto \exp\left( \frac{B_{23}^2 k}{2A_0^2(K^2+1)D_{33}}t\right),
\end{equation}
where $k$ and $D_{33}$ are positive constants. 

The exponential growth of the complementary function would render the multiple-scales expansion from Eq.~\eqref{eq:multiple-scale-expansion-phi} invalid, so we deduce by contradiction that $f(t,\tau)$ must be zero. If $f(t,\tau)$ is identically zero then so is $\phi_1^{(1)}$ due to Eq.~\eqref{app:eq:f}, and likewise $g(t,\tau)=0$ due to Eq.~\eqref{app:eq:g}.

\subsection{Solution of first-order equations}

After substituting the solvability condition that $\phi_1^{(1)}=0$ into Eq.~\eqref{app:eq:first-order-dynsys}, we find a system of equations for $x_1^{(1)}$ and $\phi_1^{(0)}$ only,
\begin{multline}
\begin{pmatrix}
1 & -\rho\sin(\phi^{(0)}_1-\xi) \\ 0 & 1
\end{pmatrix}
\begin{pmatrix}
x_{1,t}^{(1)} \\ \phi_{1,\tau}^{(0)}
\end{pmatrix} 
= 
\tilde{\mathbf{S}}_0^{-1}(\phi_1^{(0)})\begin{pmatrix} -kx_1^{(1)} \\ 0 \end{pmatrix} 
\\
- \tilde{\mathbf{S}}_0^{-1}(\phi_1^{(0)})\tilde{\mathbf{C}}_1(\ex,\phi_1^{(0)},\phi_2^{(0)})\tilde{\mathbf{S}}_0^{-1}(\phi_2^{(0)}) 
\begin{pmatrix}
-k\rho\cos(\phi^{(0)}_2-\xi) \\ T_0
\end{pmatrix}.
\label{eq:first-order-eqn-without-secular-terms}
\end{multline}
We solve these equations in the same way that we approached the leading-order system from Eq.~\eqref{eq:multiplescales-order0}, by finding series expansions of the resistance matrices with respect to the asymptotic parameter $(\pi N)^{-1} \ll 1$. 

First, we expand the cross-interaction resistance matrix in powers of $(\pi N)^{-1}$ as
\begin{equation}
    \tilde{\mathbf{C}}_1(\ex,\phi_1,\phi_2) = \hat{\mathbf{C}}_0 + \hat{\mathbf{C}}_{-1} + \hat{\mathbf{C}}_{-2} + \hat{\mathbf{C}}_{-3} + \hat{\mathbf{C}}_{-4}. \label{eq:C-series}
\end{equation}
Note that this series terminates with $\hat{\mathbf{C}}_{-4} \sim \mathcal{O}((\pi N)^{-4})$, unlike the expansion of $\tilde{\mathbf{S}}_0^{-1}$ which is infinite, because we take the inverse of a matrix with a finite expansion. 

Starting from Eqs.~\eqref{eq:changeofbasis}-\eqref{eq:cross-definition} we deduce, after some algebra, that
\begin{eqnarray}
\hat{\mathbf{C}}_0 &=& -\frac{1}{8 \pi \mu}\begin{pmatrix}
2A_0^2 & ~0~ \\ ~0 & ~0~
\end{pmatrix} \sim \mathcal{O}((\pi N)^{0}), \label{eq:matrix-Chat0}\\
\hat{\mathbf{C}}_{-1} &=& -\frac{1}{8 \pi \mu} \begin{pmatrix}
2A_0\Delta A(\cos(2\phi_1)+\cos(2\phi_2)) & ~0~ \\ ~0 & ~0~
\end{pmatrix}\sim \mathcal{O}((\pi N)^{-1}), \\
\hat{\mathbf{C}}_{-2} &=& -\frac{1}{8 \pi \mu} \begin{pmatrix} \substack{
	2\Delta A^2\cos(2\phi_1)\cos(2\phi_2) + \Delta A^2\sin(2\phi_1)\sin(2\phi_2) \\ + A_{23}^2\sin(\phi_1)\sin(\phi_2)} & {\scriptstyle -2A_0B_{23}\sin(\phi_2)-A_{23}B_{33}\sin(\phi_1)} \\ {\scriptstyle -2A_0B_{23}\sin(\phi_1)-A_{23}B_{33}\sin(\phi_2)} & B_{33}^2
\end{pmatrix}\sim \mathcal{O}((\pi N)^{-2}),\\
\hat{\mathbf{C}}_{-3} &=& -\frac{1}{8 \pi \mu} \begin{pmatrix}
0 & \substack{-2\Delta A B_{23} \cos(2\phi_1)\sin(\phi_2) \\ + \Delta A B_{23} \sin(2\phi_1)\cos(\phi_2)} \\ \substack{-2\Delta A B_{23} \cos(2\phi_2)\sin(\phi_1) \\ + \Delta A B_{23} \sin(2\phi_2)\cos(\phi_1)} & 0
\end{pmatrix}\sim \mathcal{O}((\pi N)^{-3}), \\
\hat{\mathbf{C}}_{-4} &=& -\frac{1}{8 \pi \mu} \begin{pmatrix}
~0~ & ~0~ \\ ~0 & 2 B_{23}^2\sin(\phi_1)\sin(\phi_2)+B_{23}^2\cos(\phi_1)\cos(\phi_2)
\end{pmatrix}\sim \mathcal{O}((\pi N)^{-4}). \label{eq:matrix-Chat4}
\end{eqnarray}
This expansion relies on the scaling of the coefficients $A_{ij}, B_{ij}, D_{ij}$ with respect to the dimensionless parameter $(\pi N)^{-1} \ll 1$, which can be inferred from RFT calculations \cite{PRF2021}.

Using the expansion of $\tilde{\mathbf{S}}_0^{-1}$ from Eqs.~\eqref{eq:matrix-Shat2}-\eqref{eq:matrix-Shat0} and that of $\tilde{\mathbf{C}}_1$ from Eqs.~\eqref{eq:matrix-Chat0}-\eqref{eq:matrix-Chat4}, we also find the expansion
\begin{equation}
    \tilde{\mathbf{S}}_0^{-1}(\phi_1)\tilde{\mathbf{C}}_1(\ex,\phi_1,\phi_2)\tilde{\mathbf{S}}_0^{-1}(\phi_2)  = \mathbf{M}_{2}(\phi_1,\phi_2) + \mathbf{M}_{0}(\phi_1,\phi_2) + \mathcal{O}\left((\pi N)^{-1}\right), \label{eq:SCS-series}
\end{equation}
where the first two coefficients are equal to 
\begin{eqnarray}
\mathbf{M}_{2}(\phi_1,\phi_2) &=& -\frac{1}{8 \pi \mu}\begin{pmatrix}
~0 & 0 \\ ~0 & B_{33}^{2}D_{33}^{-2}
\end{pmatrix}\sim \mathcal{O}((\pi N)^{2}), \label{eq:matrix-M2}
\\
\mathbf{M}_{0}(\phi_1,\phi_2) &=& -\frac{1}{8 \pi \mu}\begin{pmatrix}
~2 & 0 \\ ~0 & B_{23}^{2}D_{33}^{-2}\cos(\phi_1)\cos(\phi_2)
\end{pmatrix}\sim \mathcal{O}((\pi N)^{0}). \label{eq:matrix-M0}
\end{eqnarray}

By substituting the expansions from Eqs.~\eqref{eq:Sinv-series} and \eqref{eq:SCS-series} into Eq.~\eqref{eq:first-order-eqn-without-secular-terms} and inverting the matrix on the left-hand side of Eq.~\eqref{eq:first-order-eqn-without-secular-terms}, we find that
\begin{equation}
    \begin{pmatrix}
x_{1,t}^{(1)} \\ \phi_{1,\tau}^{(0)}
\end{pmatrix} \approx 
\begin{pmatrix}
~1~ & \rho\sin(\phi^{(0)}_1-\xi) \\ ~0~ & 1
\end{pmatrix}
\begin{pmatrix} -\frac{k}{A_0}x_1^{(1)}-\frac{k}{4\pi\mu}\rho\cos(\phi^{(0)}_2-\xi) \\ \frac{B_{33}^{2}T_0}{8\pi \mu D_{33}^2} -\frac{kB_{23}}{A_0D_{33}}x_1^{(1)}\sin(\phi_1^{(0)}) +  \frac{B_{23}^{2}T_0}{8\pi\mu D_{33}^2} \cos(\phi_1^{(0)})\cos(\phi_2^{(0)})
\label{eq:first-order-eqn-after-tidyingup}
\end{pmatrix}.
\end{equation}

The leading-order terms in the equation for the lateral displacement are
\begin{equation}
\frac{\partial x_1^{(1)}}{\partial t} = -\frac{k}{A_0}x_1^{(1)}-\frac{k\rho}{4\pi\mu}\cos(\phi^{(0)}_2-\xi) - \frac{B_{33}^{2}T_0\rho}{8\pi\mu D_{33}^2}\sin(\phi^{(0)}_1-\xi), \label{eq:x1-diffeqn}
\end{equation}
up to and including $\mathcal{O}((\pi N)^{0})$ terms. The general solution of this equation has the form 
\begin{equation}
x_1^{(1)}(t,\tau) = \hat{x}_1(\tau)e^{-kt/A_0} + \gamma \sin(\phi_1^{(0)}) + \delta \cos(\phi_1^{(0)}) + \zeta \sin(\phi_2^{(0)}) + \eta \cos(\phi_2^{(0)}),
\label{eq:x(1)}
\end{equation}
and is obtained by multiplying Eq.~\eqref{eq:x1-diffeqn} by an integrating factor, $\exp(kt/A_0)$, and then integrating the right-hand side by parts using the fact that $\partial \phi_j^{(0)}/\partial t \approx T_0 D_{33}^{-1}$ to leading order.

The only coefficient from the general solution, Eq.~\eqref{eq:x(1)}, that is important for synchronization is $\eta$, which comes from the integration of the term
\begin{equation}
    e^{-kt/A_0}\int^t e^{ks/A_0}\cos(\phi_2^{(0)}(s,\tau)-\xi) \mathrm{d}s = \frac{kA_0^{-1}\cos(\phi_2^{(0)}-\xi) + T_0D_{33}^{-1}\sin(\phi_2^{(0)}-\xi)}{k^2A_0^{-2} + T_0^2D_{33}^{-2}}, \label{eq:cos-exp-integral}
\end{equation}
giving us
\begin{equation}
    \eta = -\frac{k\rho}{4\pi\mu}\left(\frac{kA_0^{-1}\cos(\xi) - T_0D_{33}^{-1}\sin(\xi)}{k^2A_0^{-2} + T_0^2D_{33}^{-2}}\right).
\end{equation}
The expression for $\eta$ can be simplified using the definitions of $\rho$ and $\xi$ from Eq.~\eqref{eq:amplitude_and_lag}, and we have
\begin{equation}
\eta = \frac{B_{23}K^2}{2\pi\mu (K^2+1)^2},
\label{eq:eta}
\end{equation}
where $K = kD_{33}/(A_0 T_0)$ as defined in Eq.~\eqref{eq:defn-K}.

\subsection{Evolution of phase difference}

From Eq.~\eqref{eq:first-order-eqn-after-tidyingup} we deduce that the slow evolution of the phase difference, at leading order, is given by
\begin{equation}
\frac{\partial}{\partial \tau}\left(\phi_1^{(0)} - \phi_2^{(0)}\right) = - \frac{kB_{23}}{A_0D_{33}}(x_1^{(1)}\sin(\phi_1^{(0)})-x_2^{(1)}\sin(\phi_2^{(0)}))
\end{equation}
By substituting the general solution for $x_j^{(1)}$ from Eq.~\eqref{eq:x(1)} and averaging over the fast time scale of rotation, we recover the classical Adler equation \cite{Adler1946} for the mean phase difference between the filaments
\def\d{{\rm d}}
\begin{equation}
\frac{\d \langle\Delta\phi\rangle}{\d \tau} = - \frac{kB_{23}\eta}{A_0D_{33}} \sin(\Delta\phi) \equiv -\frac{\sin(\Delta\phi)}{\tau_\mathrm{sync}},
\label{eq:eqn-sync}
\end{equation}
where the angle brackets denote the average over the time scale of rotation, $\langle \dots \rangle = \frac{1}{t_\mathrm{rot}} \int_0^{t_\mathrm{rot}} \dots \d t$. After substituting the expression for $\eta$ from Eq.~\eqref{eq:eta} and using the conversion $t_{\mathrm{sync}} = d\tau_{\mathrm{sync}}$, we finally deduce that the time scale for rotation is 
\begin{equation}
t_{\mathrm{sync}} = \frac{2\pi\mu d D_{33}^2}{B_{23}^2T_0}\lambda(K), \qquad \lambda(K) = \frac{(K^2 + 1)^2}{K^3}.
\label{eq:tsync}
\end{equation}
The function $\lambda(K)$ and hence the time scale for synchronization are minimized at an optimum value $K^* = \sqrt{3}$, which is independent of the distance between the filaments.

\section{Computational method}
\label{sec:comp-method}

\subsection{Instantaneous hydrodynamics}

We compute the instantaneous hydrodynamic forces on the two filaments using Johnson's slender-body theory (SBT) \cite{Johnson1980}. This requires us to solve an integral equation relating the velocity along the filament centreline, $\ub(\rb_1(s))$, and the force densities along the two filaments, $\fb_1(s)$ and $\fb_2(s)$, in the form
\begin{equation}
8\pi\mu\ub(\rb_1(s)) = \mathcal{L}[\fb_1(s)] + \mathcal{K}[\fb_1(s')] + \mathcal{J}[\fb_2(s'),\db],
\label{chp6:eq:COMP-method}
\end{equation}
where the first operator represents local effects
\begin{equation}
\mathcal{L}[\fb_1(s)] = \left[2\left(\ln\left(\frac{2}{\epsilon}\right)+\frac{1}{2}\right)\Ib + 2\left(\ln\left(\frac{2}{\epsilon}\right)-\frac{3}{2}\right)\tb_1(s)\tb_1(s)\right]\cdot \fb_1(s),
\label{chp6:eq:COMP-local}
\end{equation}
and is defined in terms of the unit tangent along the centreline, $\tb_1(s)$, and the dimensionless cross-sectional radius of the filament, $\epsilon=2r_\epsilon/L$. 
The second operator represents non-local effects
\begin{eqnarray}
\mathcal{K}[\fb_1(s')] &=& \sdint{\left[\frac{\Ib+\hat{\Rb}_0(s,s')\hat{\Rb}_0(s,s')}{|\Rb_0(s,s')|}-\frac{\Ib+\tb_1(s)\tb_1(s)}{|s'-s|}\right]\cdot \fb_1(s')} \nonumber\\ 
&+& \left(\Ib+\tb_1(s)\tb_1(s)\right)\cdot\sdint{\frac{\fb_1(s')-\fb_1(s)}{|s'-s|}},
\label{chp6:eq:COMP-nonlocal}
\end{eqnarray}
where $\Rb_0(s,s') = \rb_1(s)-\rb_1(s')$ and $\hat{\Rb}_0 = \Rb_0/|\Rb_0|$. Finally, the third operator represents interactions between the two filaments, as previously modelled by Tornberg and Shelley \cite{Tornberg2004},
\begin{equation}
\mathcal{J}[\fb_2(s'),\db] = \sdint{\left[\frac{\Ib+\hat{\Rb}_d(s,s')\hat{\Rb}_d(s,s')}{|\Rb_d(s,s')|}  + \frac{\epsilon^2}{2}\frac{\Ib-3\hat{\Rb}_d(s,s')\hat{\Rb}_d(s,s')}{|\Rb_d(s,s')|^3}\right]\cdot\fb_2(s')},
\label{chp6:eq:COMP-interaction}
\end{equation}
where $\Rb_d(s,s') = \db +\rb_2(s')-\rb_1(s)$, $\hat{\Rb}_d = \Rb_d/|\Rb_d|$ and $\db$ is the distance between the reference positions of the two filaments. The two terms in the interaction operator, $\mathcal{J}$, represent Stokeslet and source dipole contributions, respectively. 
%Note that, for the source dipole term, we have used the same prefactor of $1/2$ as in Ref.~\cite{Tornberg2004}, while a more recent study based on the Rotne-Prager-Yamakawa kernel and matched asymptotics uses a larger prefactor of $e^3/24$ \cite{Maxian2021}. \color[red]{[Do we need to meention the prefactor? This is not a technical chapter...}

We solve Eq.~\eqref{chp6:eq:COMP-method} numerically using a Galerkin method, whereby we decompose the velocities, $\ub(\rb_j(s))$, and force densities, $\fb_j(s)$, into Legendre polynomial modes. The system is then truncated to a finite number of modes \footnote{Through self-convergence tests of the numerical solution, we determined that a truncation level of $N_{\mathrm{Legendre}}=15$ Legendre polynomial modes was sufficiently accurate for the purposes of our simulations.} and inverted numerically to find the net forces and torques exerted by the filaments, $(\Fb_1,\Tb_1;\Fb_2,\Tb_2)$, for any given rigid body motion of the filaments, $(\Ub_1,\Omb_1;\Ub_2,\Omb_2)$. This is equivalent to computing the extended resistance matrix for two interacting rigid filaments, as described in previous work \cite{PRF2021}. 

\subsection{Integration in time}

To obtain the time evolution of the rotating filaments, we solve the dynamical system
\begin{equation}
    \dot{\Xb} = \Rb(\Xb)^{-1}\Fb(\Xb),
\label{chp6:eq:COMP-time-integration}
\end{equation}
where $\Xb$ describes the generalized configuration of the two filaments (their positions and orientations), while $\Fb$ encapsulates the overall dynamics (the forces and torques on both filaments) and may depend on the instantaneous configuration. The linear relationship between the two is provided by the extended resistance matrix, $\Rb(\Xb)$, which also depends on the instantaneous configuration of the two filaments and is calculated from Eqs.~\eqref{chp6:eq:COMP-method}-\eqref{chp6:eq:COMP-interaction}, as described above. 

We integrate Eq.~\eqref{chp6:eq:COMP-time-integration} in time using the classical Runge–Kutta method (RK4) within the dynamical conditions $\Fb(\Xb)$ prescribed by our minimal model for synchronization. For the case where the filament axes remain vertical, we impose the condition that $\dot{\Xb}=\mathbf{0}$ in all components except for the four degrees of freedom of the system, $x_{1,2}$ and $\phi_{1,2}$. In these four degrees of freedom, we prescribe a constant driving torque, $T_0\ez$, and an elastic restoring force, $-kx_j\ex$, for each filament. 

The initial conditions for the dependent variables are
\begin{equation}
    \phi_1(0) = 0, \quad \phi_2(0) = \frac{\pi}{2}, \quad x_1(0) = \rho\cos(\xi), \quad x_2(0) = \rho\sin(\xi), \label{chp6:eq:ICs}
\end{equation}
where the conditions for $x_{1,2}$ are informed by our knowledge of the intrinsic dependence of the lateral displacement on the phase angle, Eq.~\eqref{eq:xlagsbehindphi}. The only exceptions are the numerical simulation shown in Fig.~2a, where the phase difference $\Delta\phi = \phi_2 - \phi_1$ starts close to $\pi$, and those shown in Fig.~\ref{fig:S4-torquediff}, where the initial phase difference takes several values between $0$ and $\pi$.

In order to capture the dynamics on the fast time scale of rotation faithfully, we choose a time step smaller than the period of rotation, $t_{\mathrm{step}} = t_{\mathrm{rot}}/20$, where the intrinsic time scale for rotation $t_{\mathrm{rot}} = D_{33}/T_0$ is estimated from the SBT resistance matrix of a single filament (derived by solving Eq.~\eqref{chp6:eq:COMP-method} without the interaction term, $\mathcal{J}$). The only exception is the long-time simulation shown in Fig.~2a, where we use a larger time step, $t_{\mathrm{step}} = t_{\mathrm{rot}}/10$. 

\subsection{Extracting the time scale for synchronization}

To reduce computation time, we do not generate the full evolution of the phase difference for every set of parameters, since we know that it follows the dynamics of Adler's equation (see Fig.~2a). We run targeted simulations starting from $\Delta\phi = \pi/2$ with a total integration time of ten periods of rotation (equivalent to the inset panel on Fig.~2a). From each of these simulations we extract the time scale for synchronization by applying a linear fit to the average phase difference. The slope of the linear fit is equivalent to the rate of synchronization since expanding the solution to Adler's equation, $\langle\Delta\phi\rangle = 2\tan^{-1}\left[\tan(\Delta\phi_0/2)\exp(-(\tau-\tau_0)/\tau_{\mathrm{sync}})\right]$, near its inflection point $(\tau_0,\pi/2)$ gives
\begin{equation}
    \Delta\phi(\tau) \approx \frac{\pi}{2} - \frac{\tau-\tau_0}{\tau_{\mathrm{sync}}}.
\end{equation}
Each targeted simulation corresponds to one data point in Figs.~2b and 2c of the main manuscript, and Figs.~\ref{fig:S1-extended-range-d}, \ref{fig:S2} and \ref{fig:S4-torquediff} of the supplementary material.

\color{black}

\section{Supplementary simulations}
\label{sec:simulations}

\subsection{Extended range of interflagellar separation}

\begin{figure}[t]
    \centering
    \includegraphics[width=\textwidth]{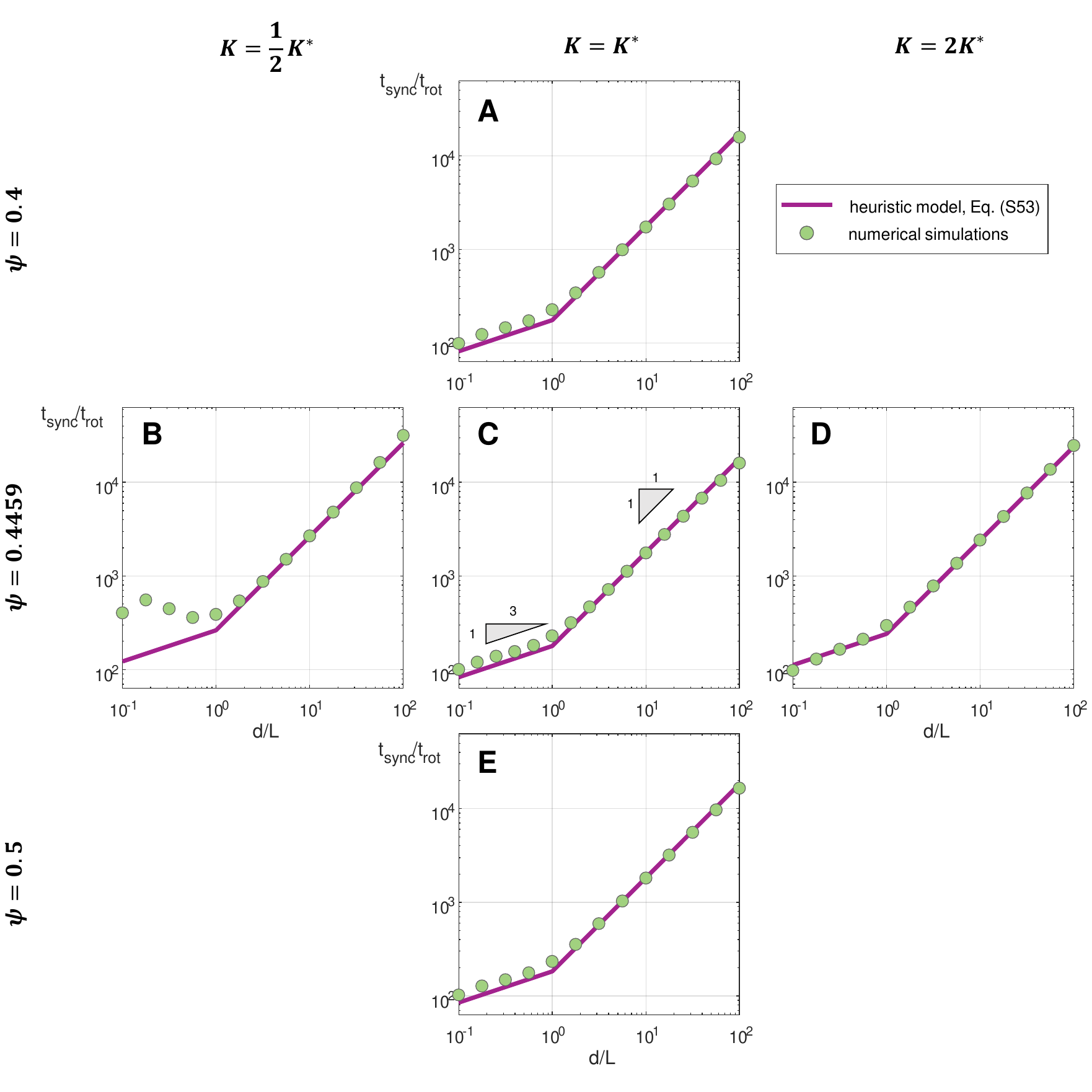}
    \caption{\textbf{Extended range of inter-flagellar distance}. The different panels show the dependence of the synchronization time on inter-filament distance for several values of the elastic tethering strength, $K$, and helix pitch angle, $\psi$, which are indicated on the margins of the figure. For $K\geq K^*$, the sharp transition between the power laws $t_\mathrm{sync}\sim (d/L)^{1/3}$ and $t_\mathrm{sync}\sim (d/L)$ at $d=L$ allows us to model the time scale of synchronization using the heuristic expression from Eq.~\eqref{eq:heuristic_model}, where the parameter $\alpha$ is identified directly by our far-field theory, Eq.~\eqref{eq:tsync}.}
    \label{fig:S1-extended-range-d}
\end{figure}

We performed additional simulations to extend the range of interflagellar separation below one filament length, down to $d/L = 0.1$, which is comparable to the width of a bacterial cell body. The results of the simulations are shown in Fig.~\ref{fig:S1-extended-range-d}. We observe that the scaling $t_\mathrm{sync}\sim d/L$ derived in our analytical far-field theory changes to a scaling $t_\mathrm{sync}\sim (d/L)^{1/3}$ when the distance between the filaments is smaller than one filament length, but only when the strength of the elastic restoring force, $K$, is sufficiently large (compare panel B of Fig.~\ref{fig:S1-extended-range-d} with panels C and D). 

This observation suggests two ways in which our far-field theory of synchronization contributes to our understanding of real bacterial flagellar synchronization. Firstly, by identifying the value of elastic compliance that is optimal for far-field synchronization, our theory helps to distinguish between the different regimes of synchronization in the near field. Secondly, in the regime where the $(d/L)^{1/3}$ scaling holds, we may use our far-field theory to get the order of magnitude of the time scale of synchronization in the near-field as well. Based on the observation that the transition between the two power laws occurs sharply around $d=L$, we suggest the following heuristic model for the time scale of synchronization
\begin{equation}
    t_\mathrm{sync} = \begin{cases}
    \alpha \left(d/L\right)^{1/3},& \text{if } d< L\\
    \alpha d/L,      & \text{if } d\geq L
\end{cases}
\label{eq:heuristic_model}
\end{equation}
where we have an exact expression for the constant $\alpha$ from our far-field theory (no fitting parameters, calculated directly from Eq.~\eqref{eq:tsync}). This approach provides reasonable agreement with numerical simulations, as shown in Fig.~\ref{fig:S1-extended-range-d}.

\newpage
\subsection{Higher number of kinematic degrees of freedom}

\begin{figure}[t]
    \centering
    \includegraphics[width=\textwidth]{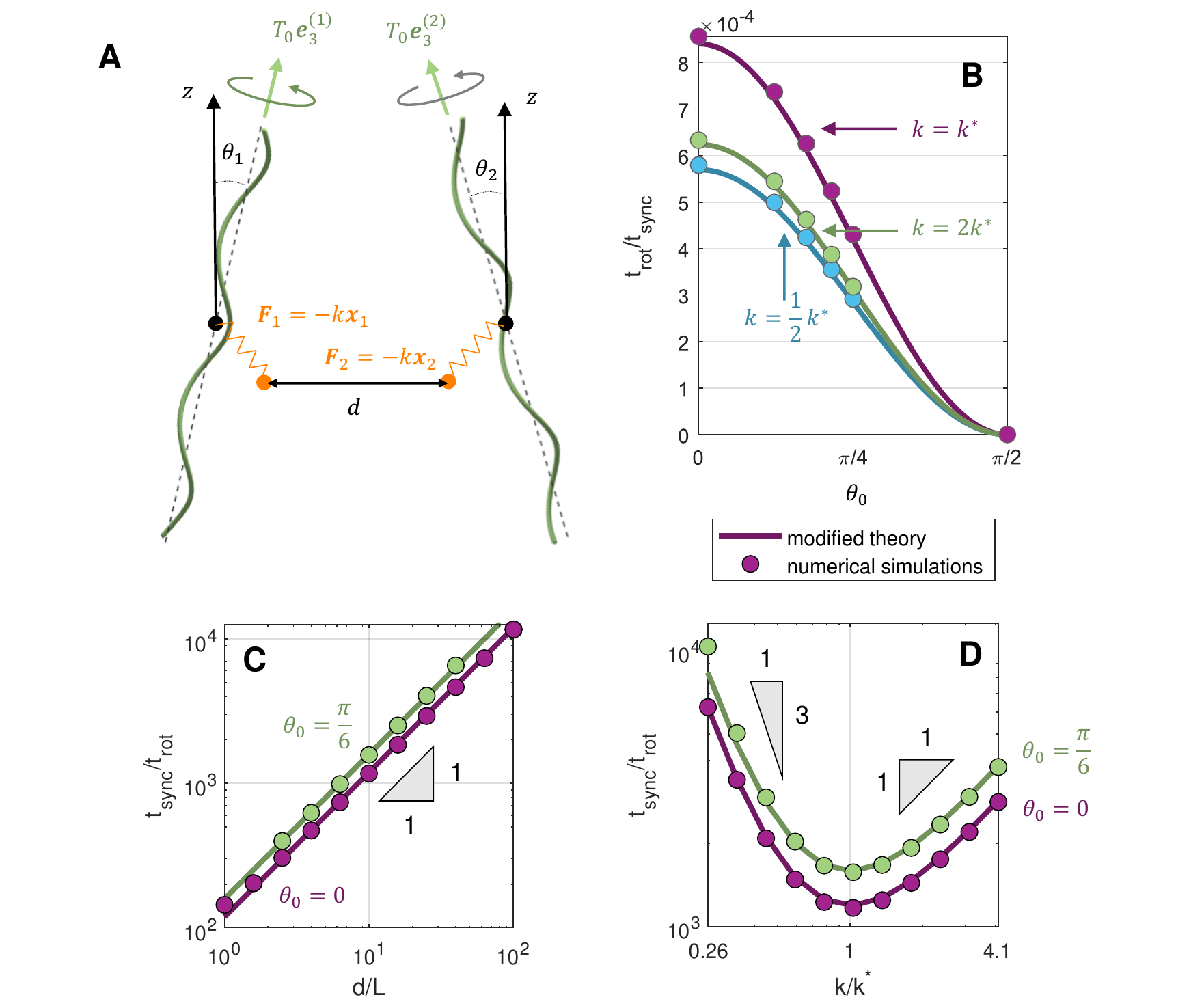}
    \caption{\textbf{Higher number of kinematic degrees of freedom}. (A) Sketch of the setup with two filaments rotating and translating freely (i.e.~no restrictions on the possible orientation or position of the filaments) under the action of a constant driving torque along the central axis of each filament, $T_0\eee$, and an elastic restoring force $-k\xb$ acting to bring back the midpoint of the axis of rotation (black dot) back to a reference position (orange dot). The inclination angles $\theta_1(t), \theta_2(t)$ vary with time. (B) Rate of synchronization of two freely rotating filaments against the initial inclination angle, $\theta_1(0) = -\theta_2(0) = \theta_0$. (C) Time scale of synchronization for two freely rotating filaments against the inter-filament distance. (D) Time scale of synchronization for two freely rotating filaments against the elastic compliance.  (B-D) The modified theory consists of multiplying the time scale of synchronization from Eq.~\eqref{eq:tsync} by a factor of $\frac{2}{3}\sec^2\theta_0$. Hence, the rate of synchronization in panel B is proportional to $\cos^2\theta_0$.}
    \label{fig:S2}
\end{figure}

In Section \ref{sec:theory}, we restricted the number of degrees of freedom to two for each filament (one translational and one rotational degree of freedom) in order to make the analytical calculations tractable. We perform additional numerical simulations in this section, where we relax the constraints that the axes of rotation of the filaments remain vertical (i.e. $\eee = \ez$) and oscillate in {the $x$ direction only (i.e. $\dot{y}=\dot{z}=0$)}. We allow both the orientation and position of the filament to vary freely, and we prescribe a constant driving torque $\Tb = T_0 \eee$ about the axis that goes through the centre of each helical filament ($\eee$, which may now deviate from the vertical, $\ez$). We also impose a general elastic restoring force $\Fb = -k\xb$, where the vector $\xb$ gives the displacement of the midpoint of the axis of rotation relative to a reference position. Since the filament axis is allowed to deviate from the vertical, we measure this inclination by the angle $\theta$, where $\eee \cdot \ez = \cos\theta$. This new setup is illustrated in Fig.~\ref{fig:S2} A. 

Because the torque is applied on the long axis of the filaments, which are assumed to be long and slender (making the resistance to rotation about the long axis much smaller than the resistance to rotation about the two minor axes), we find in numerical simulations that the filaments rotate stably around the long axis with only small oscillations in the angle $\theta(t)$ as a function of time (results not shown). Hence, the initial conditions imposed on the inclination angle is a reliable indicator of the configuration in which the filaments continue to rotate as they synchronize.

The initial conditions we prescribe on the filaments are that their axes are pointing symmetrically towards each other, such that $\theta_1(0) = \theta_0$ and $\theta_2(0) = - \theta_0$, see Fig.~\ref{fig:S2} A. Hence, the angle between the two axes is $2\theta_0$. In Fig.~\ref{fig:S2} B, the filaments are inclined at various angles from $\theta_0 = 0$ to $\theta_0 = \pi/2$, while we keep a fixed distance $d/L =10$ between the filaments and consider three special values of elastic compliance (the far-field optimum, above the optimum, and below the optimum). We empirically find that our result from Eq.~\eqref{eq:tsync} can be multiplied by a factor $\frac{2}{3}\sec^2(\theta_0)$ to reproduce the time scales of synchronization observed in numerical simulations. The numerical factor of $2/3$ arises purely from changing the number of degrees of freedom in our model, and is independent of the parameters in our model. This is demonstrated by the fact that the modified theory matches perfectly with numerical simulations as we vary the distance between the filaments (Fig.~\ref{fig:S2} C), the elastic compliance (Fig.~\ref{fig:S2} D), and the geometry of the filaments (we have tested the normal and the semicoiled geometry of bacterial flagella, results not shown). Crucially, the inclination of the filaments only multiplies the typical time scale for synchronization and does not affect the optimum elastic compliance for synchronization, which is still given by the global minimum of the function $\lambda(K)$ from Eq.~\eqref{eq:tsync}.
%from which we draw the conclusion that filaments rotating anti-parallel to each other (driving torques pointing towards each other, with an angle $2\theta_0 = \pi$ between the axes) do not synchronize. 

\subsection{Non-identical flagella}

In Section \ref{sec:leadingorderkinematics}, we showed that in the absence of hydrodynamic interactions (in the limit $d\to\infty$) each helical filament rotates with intrinsic angular velocity $\Omega_0 = T_0 D_{33}^{-1}$. If the two flagella are not identical, but have a small mismatch between their intrinsic rotation rates, i.e.~$\Delta\Omega_0 = \Omega_0^{(2)} - \Omega_0^{(1)} = \mathcal{O}(d^{-1})$, this discrepancy will appear in the equations of motion for the two flagella at first-order in $d^{-1}$. Hence, the Adler equation for synchronization from Eq.~\eqref{eq:eqn-sync} can be adapted to allow for non-identical flagella, taking the form
\begin{equation}
\frac{\d \langle\Delta\phi\rangle}{\d t} = \Delta\Omega_0 -\frac{\sin(\Delta\phi)}{t_\mathrm{sync}}.
\label{eq:eqn-sync-nonidentical}
\end{equation}
The right-hand side of this equation tells us that the hydrodynamic interactions between the filaments may counterbalance the mismatch in intrinsic rotation rates if the mismatch is not larger than $\mathcal{O}(d^{-1})$, the size of hydrodynamic effects. If $|\Delta\Omega_0| \leq t_\mathrm{sync}^{-1}$, the phase difference between the two flagella tends to an equilibrium value
\begin{equation}
    \Delta\phi_\infty = \arcsin\left(\Delta\Omega_0t_\mathrm{sync}\right),
    \label{eq:steadystates_nonidentical}
\end{equation}
meaning that two non-identical rotating helices ($\Delta\Omega_0 \neq 0 $) phase-lock to a finite phase difference. 

\begin{figure}
    \centering
    \includegraphics[width=\textwidth]{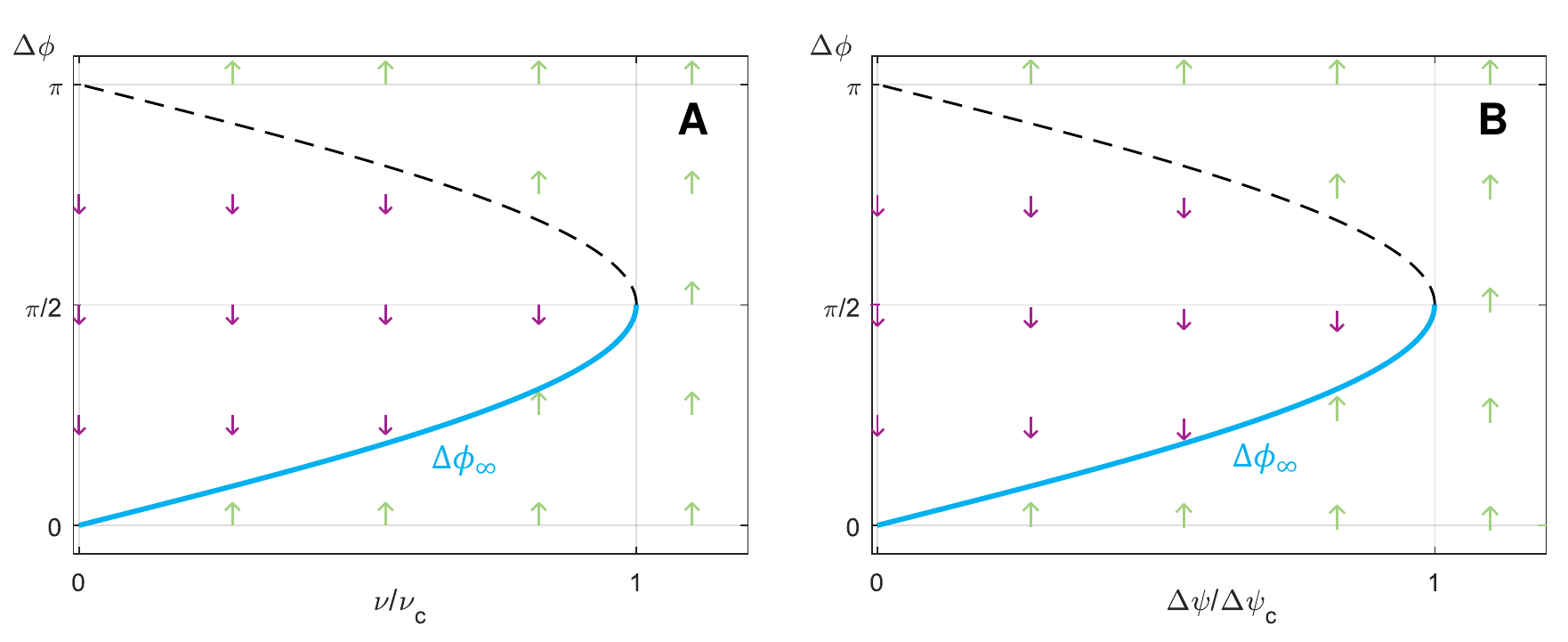}
        \caption{\textbf{synchronization of non-identical flagella}. A small mismatch in (A) the driving torques, $(1 \pm \nu)T_0$,  or (B) the pitch angle of the helices, $\psi \pm \Delta\psi$,  can induce a finite equilibrium phase difference between the rotating filaments. The bifurcation diagrams show a saddle-node bifurcation at the critical values $\nu = \nu_c$ and $\Delta\psi = \Delta\psi_c$ (numerical values provided in Table \ref{table:SM}), where the stable branch (continuous blue line) coalesces with the unstable branch (dashed black line). The lines are theoretical predictions (from Eqs.~\eqref{eq:steadystates_nonidentical}-\eqref{eq:critical_dpsi}) while the arrows indicate the rate of change of the phase difference (positive or negative) measured in numerical simulations.}
    \label{fig:S4-torquediff}
\end{figure}

A concrete example of inhomogeneity that leads to a mismatch in intrinsic rotation rates is when the two flagella are driven by constant but not identical torques, $(1\pm \nu)T_0$, where the fractional torque difference $2\nu = \mathcal{O}(d^{-1})$ is small. In  this case, the mismatch in intrinsic rotation rates is simply $\Delta\Omega_0 = 2\nu T_0D_{33}^{-1}$ and the filaments phase-lock to a finite phase difference so long as
\begin{equation}
    |\nu| \leq \nu_c = \frac{D_{33}}{2T_0t_\mathrm{sync}}. \label{eq:critical_nu}
\end{equation}
If we follow the evolution of the steady states of Eq.~\eqref{eq:eqn-sync-nonidentical} as we increase the fractional torque difference, $\nu$, we obtain a bifurcation diagram such as the one in Fig.~\ref{fig:S4-torquediff} A (results consistent with Ref.~\cite{Reichert2005}). The theory predicts that the stable solution (solid blue line) collides with the unstable solution (dashed black line) through a saddle-node bifurcation at the critical value $\nu=\nu_c$ , after which the system no longer has any steady states and the phase difference between the filaments drifts continuously due to the mismatch in intrinsic rotation rates. This theoretical prediction is confirmed by numerical simulations. The arrows in Fig.~\ref{fig:S4-torquediff} A indicate the sign of $\d\Delta\phi/\d t$ for various values of $\nu$ and $\Delta\phi$, determined from numerical simulations.

Alternatively, the mismatch in intrinsic rotation rates can arise from differences in the geometry of the filaments. For instance, small perturbations to the pitch angle of the helices, $\psi \pm \Delta\psi$, make the filaments phase-lock to a finite phase difference so long as
\begin{equation}
    |\Delta\psi| \leq \Delta\psi_c = \left(2\frac{\partial \Delta\Omega_0}{\partial\psi}t_\mathrm{sync}\right)^{-1}, \label{eq:critical_dpsi}
\end{equation}
which is illustrated in the bifurcation diagram of Fig.~\ref{fig:S4-torquediff} B. To determine the stable and unstable branches more accurately, we calculated $\partial\Delta\Omega_0/\partial\psi$ numerically using slender-body theory computations on individual filaments, although the derivative can also be derived analytically since $\Delta\Omega_0(\psi) = T_0/D_{33}(\psi)$ and we have closed analytical expressions for $D_{33}(\psi)$ from resistive-force theory \cite{PRF2021}. 

\section{Simulation parameters}
\label{sec:parameters}

In this section, we list all the simulation parameters used to generate the results from the main manuscript, Table \ref{table:main}, and the supplementary material, Table \ref{table:SM}. We also provide some clarifications regarding the definition of the dimensionless elastic compliance, pitch angle and filament thickness.

The parameters are non-dimensionalised such that the contour length of the filament is $L=2$, and the driving torque is $T_0 = 1$. Therefore, the dimensionless elastic compliance is given by $k = \tilde{k}\tilde{L}^2/4\tilde{T}_0$, where tildes denote the dimensional counterpart of each quantity. In terms of the helical amplitude, $r$, and helical pitch, $p$, the pitch angle of the helix is defined as $\psi = \tan^{-1}(2\pi r/p)$. The dimensionless filament thickness is defined as $\epsilon = 2\tilde{r}_\epsilon/\tilde{L}$, where $\tilde{r}_\epsilon$ is the radius of the cross-section of the filament. 

\begin{table}[H]

\begin{ruledtabular}
\begin{tabular}{l c c c}
\textbf{Input parameter} & \textbf{Fig.~2a} & \textbf{Fig.~2b} & \textbf{Fig.~2c}\\
\hline
Distance, $d/L$ & 2 & 1 -- 100 & 10\\
Torque, $T_0$ & 1 & 1 & 1 \\
Elastic compliance, $k$ & 303.5 & 397.9 & 99.47 -- 1592\\
Pitch angle, $\psi$ & 0.504 & 0.446 & 0.446\\
Number of helical turns, $N$ & 2.5 & 2.5 & 2.5\\
Filament thickness, $\epsilon$ & 0.00379 & 0.00377 & 0.00377\\
\toprule
\textbf{Inferred parameter} & \textbf{Fig.~2a} & \textbf{Fig.~2b} & \textbf{Fig.~2c}\\
\hline
Optimal elastic compliance, $k^*$ & 312.1 & 386.5 & 386.5
\end{tabular}
\end{ruledtabular}
\caption{Simulation parameters for the figures in the main manuscript.}
\label{table:main}
\end{table}

\begin{table}[H]

\begin{ruledtabular}
\begin{tabular}{l c c | c c c | c c }
\textbf{Input parameter} & \multicolumn{2}{c}{\textbf{Fig.~S1}} & \multicolumn{3}{c}{\textbf{Fig.~S2}} & \multicolumn{2}{c}{\textbf{Fig.~S3}} \\
 & \textbf{A, C, E} & \textbf{B, C, D} & \textbf{B} & \textbf{C} & \textbf{D} & \textbf{A} & \textbf{B} \\
 \hline
Distance, $d/L$ & 0.1 -- 100 & 0.1 -- 100 & 10 & 1 -- 100 & 10 & 10 & 10 \\
Torque, $T_0$ & 1 & 1 & 1 & 1 & 1 & $1\pm \nu$ & 1 \\
Elastic compliance, $k$ & 397.9 & 198.9 -- 795.8 & 198.9 -- 795.8 & 397.9 & 99.47 -- 1592 & 397.9 & 397.9 \\
Pitch angle, $\psi$ & 0.4 -- 0.5 & 0.4459 & 0.4459 & 0.4459 & 0.4459 & 0.4459 & $0.4459 \pm \Delta\psi$\\
Number of helical turns, $N$ & 2.5 & 2.5 & 2.5 & 2.5 & 2.5 & 2.5 & 2.5 \\
Filament thickness, $\epsilon$ & 0.00377 & 0.00377 & 0.00377 & 0.00377 & 0.00377 & 0.00377 & 0.00377 \\
\toprule
\textbf{Inferred parameter} & \multicolumn{2}{c}{\textbf{Fig.~S1}} &\multicolumn{3}{c}{\textbf{Fig.~S2}} & \multicolumn{2}{c}{\textbf{Fig.~S3}} \\
 & \textbf{A, C, E} & \textbf{B, C, D} & \textbf{B} & \textbf{C} & \textbf{D} & \textbf{A} & \textbf{B} \\
 \hline
Optimal elastic compliance, $k^*$ & 316.9 -- 468.8 & 386.5 & 386.5 & 386.5 & 386.5 & 386.5 & 386.5 \\
Critical torque difference, $\nu_c$ & -- & -- & -- & -- & -- & $4.460 \times 10^{-5}$ & -- \\
Critical pitch angle difference, $\Delta\psi_c$ & -- & -- & -- & -- & -- & -- & $1.204 \times 10^{-5}$\\
\end{tabular}
\end{ruledtabular}
\caption{Simulation parameters for the figures in the supplementary material.}
\label{table:SM}
\end{table}